\documentclass{nature}
\usepackage{amsmath}
\usepackage{graphicx}
\usepackage{verbatim}
\usepackage{media9}
\usepackage{longtable}
\graphicspath{ {./images/} }

\title{Patchy nuclear chain reactions}

\author{E.~Dumonteil$^{1,2}$\thanks{Corresponding author.} \and R.~Bahran$^3$ \and T.~Cutler$^3$ \and B.~Dechenaux$^1$ \and T.~Grove$^3$ \and J.~Hutchinson$^3$ \and G.~McKenzie$^3$ \and A.~McSpaden$^3$ \and W.~Monange$^1$ \and M.~Nelson$^3$ \and N.~Thompson$^3$ \and A.~Zoia$^4$}

\date{}

\begin{document}

\maketitle

\begin{affiliations}
 \item Institut de Radioprotection et de S\^uret\'e Nucl\'eaire, PSN-EXP/SNC/LN, 92262 Fontenay-aux-Roses, France
 \item Institut de Recherche sur les Lois Fondamentales de l'Univers, CEA/DRF/IRFU/DPhN/LEARN, 91191 Gif/Yvette, France
 \item Los Alamos National Laboratory, NEN-2 Group, Los Alamos, NM 87545, United States
 \item DES-Service d'\'etudes des r\'eacteurs et de math\'ematiques appliqu\'ees (SERMA), CEA, Universit\'e Paris-Saclay, F-91191, Gif-sur-Yvette, France
\end{affiliations}

\begin{abstract}

Stochastic fluctuations of the neutron population within a nuclear reactor are typically prevented by operating the core at a sufficient power, since a deterministic behavior of the neutron population is required by automatic safety systems to detect unwanted power excursions. 
Recent works however pointed out that, under specific circumstances, non-Poissonian patterns could affect neutron spatial distributions. This motivated an international program to experimentally detect and characterize such fluctuations and correlations, which took place in 2017 at the Rensselaer Polytechnic Institute Reactor Critical Facility.
The main findings of this program will indeed unveil patchiness in snapshots of neutron spatial distributions -obtained with a dedicated numerical twin of the reactor- that support this first experimental characterization of the 'neutron clustering' phenomenon, while a stochastic model based on reaction-diffusion processes and branching random walks will reveal the key role played by the reactor intrinsic sources in understanding neutron spatial correlations.

\end{abstract}

\section{Introduction}

A collection of independent individuals that move, reproduce and die may undergo, under certain circumstances, wild fluctuations at the local and global scales, inducing a strongly non-Poissonian spatial patchiness: this phenomenon, dubbed `clustering', has been often observed in the context of life sciences, including the spread of epidemics \cite{sunPLR, liAMC, dumonteilPNAS}, the growth of bacteria on Petri dishes \cite{houchPRL, houchPRE1}, the dynamics of ecological communities \cite{bailey, youngNATURE}, and the mutation propagation of genes \cite{dawson,cox}. The key ingredient behind clustering is the asymmetry between death occurring everywhere and birth being only possible close to a parent particle; particle diffusion has a smoothing effect on the wild spatial patterns induced by the parent-child correlations \cite{zhangJSTAT}. The occurrence of clustering is the signature of strong deviations from the expected behavior of such systems, and has been shown to be enhanced in low-dimensional systems ($d=1$ or $d=2$), especially when the population is fairly diluted: a deterministic description would be thus meaningless, since the typical size of the fluctuations might be of the same order of magnitude as the average particle density \cite{houchPRE2, houchPRE3}.
The evolution of the neutron population in a nuclear reactor is also subject to random displacements (diffusion), births (fission events on heavy nuclei leading to secondary neutrons), and deaths (capture events on nuclei leading to the disappearance of the colliding neutrons). Therefore, it has been suggested that clustering might occur in both experiments and numerical simulations involving nuclear reactors operated at low power, i.e., low neutron density \cite{dumonteilANNALS, zoiaPRE}, or in spent-fuel pools.
Understanding space-time fluctuations of the neutron population plays a central role for nuclear safety, especially in connection with reactor control at start-up and shutdown.
Indeed, the main paradigm relies on the assumption that the transition from the stochastic to the deterministic regime is well defined and only relies only on the power of the system (directly related to the number of neutrons). Operating the reactor in the deterministic regime ensures that power variations can be related to a change of the state of the reactor that requires proper human/automatic action. For instance, a sudden raise in the reactor power can supposedly sign incidental/accidental conditions, and demands to start a reactor tripping procedure (initiated by automatic protection systems of most commercial reactors). The persistence of stochastic effects at operating conditions (in particular during the startup phase) might instead screen this state of the reactor and shall therefore be avoided. In this respect, the appearance of fission-induced correlations in nuclear systems has been extensively investigated \cite{bellBOOK, natelson, williamsBOOK, PazsitBOOK}; furthermore, neutron clustering has drawn much attention in recent years, with a special emphasis on numerical (Monte Carlo) simulations \cite{dumonteilANNALS, zoiaPRE, demulatierJSTAT, suttonNET, brownOECD, nowakANNALS, miaoANNALS}. 
Quantifying and characterizing fluctuations and spatial correlations in a low-power reactor has a theoretical and practical interest: the experimental validation of theoretical fluctuations and correlations models at low power would allow to scale them at higher powers, numericaly far from reach.  This interest motivated an international collaboration gathering LANL, IRSN and CEA, with three objectives: designing ad-hoc experiments and dedicated detectors to extract information on neutron fluctuations; building a numerical twin of the operating reactor so as to support experimental results and to fill the gaps of experimental measurements; and deriving a proper mathematical framework to interpret the obtained results. The experiments took place at the Reactor Critical Facility (RCF)~\cite{CaSPER} of the Rensselaer Polytechnic Institute (RPI) over a week in August 2017 and were followed by the analysis of measured data and their numerical reproduction. The main observables were fluctuations and spatial correlations in the neutron population, as well as their dependence on reactor power. In this work, RCF experimental data will be contrasted to Monte Carlo simulations performed by using the MORET6 code\footnote{MORET6 is based on MORET5~\cite{moret5} and features specific correlated physics functionalities described in Supplementary Material D}: we will illustrate the difficulties to characterize a clear transition from the deterministic to the stochastic regime, while unravelling striking temporal and spatial features of the neutron population within the reactor core during operation. We will noticeably point out, whenever observed on short time scales, that the chain reaction within the core present a "blinking" temporal behavior and "clustered" spatial distributions of neutrons, and we will build a stochastic model of the nuclear chain reaction based on reaction-diffusion processes and branching random walks a meme to finely reproduce the quantitative behavior of noise and spatial correlations observed during the experiments.

\section{Experimental setup and numerical experiments} \label{sec:experiment}
The Walthousen Reactor Critical Facility (RCF) at the Rensselaer Polytechnic Institute (RPI) is a zero-power education, research, and training reactor operated with 4.81\% enriched UO$_2$ ceramic fuel with stainless steel~\cite{CaSPER}. The fuel has an active length of 36 inches and the fuel pins are arranged in a square lattice with a pitch of 0.64 inches. The reactor is licensed to operate up to 100~Watts. Due to the low burnup, the fuel is essentially fresh. The core is housed in an open pool tank. This type of reactor is a good choice for performing the sought measurements, for several reasons. First, the ability to measure low neutron densities is required to meet the measurement goals. This therefore demands a low power level. Due to the absence of noticeable burnup, the fuel inside zero-power reactors is typically very well characterized, contrary to fuel from reactors with significant burnup, and allows entering the core for direct manipulation of experimental equipment. These considerations have been documented previously~\cite{CaSPER}. In addition, the RCF core is sufficiently large for spatial effects to be relevant for the purpose of the experimental campaign, as supported by preliminary simulations.

\begin{figure}[h]
    \centering
    \includegraphics[width=13cm, height=15cm]{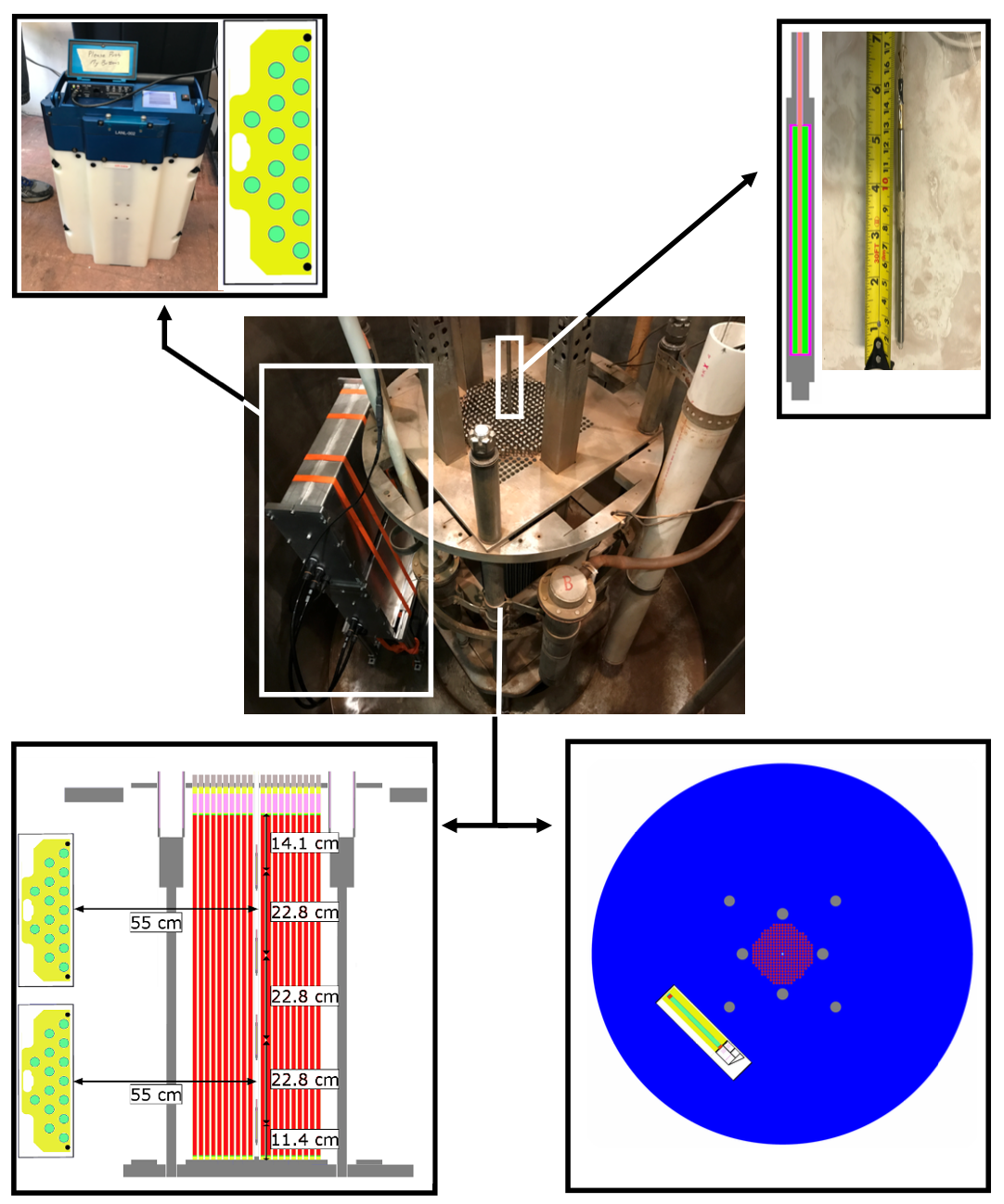}
    \caption{Experimental setup of the RCF experiments to characterize fluctuations and correlations. Central picture: top view of the RCF nuclear reactor, with 2 NOMAD detectors in their water-proof cases on the left side. The NOMAD detectors are made of 15 $^3$He tubes (top left image: picture and MORET6 modeling of a NOMAD detector out of its case). A pin-cell equipped with 4 $^3$He tubes (top right image) has been placed in the center of the reactor. Both the reactor and the detectors were modeled using MORET6 (bottom left image: side view, bottom right image: top view).}
    \label{fig:1_All_detectors2}
\end{figure}

The detection system used during the experiments was composed of three kinds of detectors. Multiple uncompensated ion chambers and BF$_3$ detectors were used for measuring reactor power. As shown by Figure~\ref{fig:1_All_detectors2}, where all detection systems are pictured, one of the central fuel pins was equipped with four small $^3$He detectors. These in-core detectors allowed to monitor the neutron spatial distribution and spatial correlations at very low power, thanks to their small saturation threshold. To finely reconstruct these observables at higher power, two LANL Neutron Multiplicity $^3$He Array Detector (NoMAD) systems were placed on the side of the reactor core. Each NoMAD system contains 15 $^3$He tubes embedded in polyethylene and arranged into three rows. Each tube is 15 inches in length \cite{SCRaP}; additional information on the $^3$He tubes is given in Table 1. Details about the reactor and the detection setup are given in Supplementary Material Section E.

To support both the design of this experiment and its analysis, special functionalities were developed within the MORET6 code (see Supplementary Material Section D) so as to numerically reproduce both the reactor, the detection system and the nuclear chain reaction. Indeed, estimating fluctuations as well as spatial and temporal correlations can be performed using the so-called "analog neutron transport" \cite{petitAnalog}: the unprecedented objective was to lean on massively parallel computing capabilities to explicitly simulate every single neutron of this sub-Watt nuclear device. A coupling with the LLNL Fission Library ~\cite{Freya} enables the emission of completely correlated fission secondaries from individual realizations of fission processes on an event-by-event basis, up to 20 MeV, for $^{235}$U (1.5\% of the fuel mass fraction) and $^{238}$U (30\% of the fuel mass fraction) in the case of neutron-induced fissions, relying on measurements of neutron multiplicity distributions made by Zucker and Holden~\cite{Zucker}. In the case of spontaneous fissions, the full multiplicity distribution data of $^{238}$U (the main contributor for spontaneous fissions in the RCF reactor) are also provided by Holden~\cite{Zucker2} while it uses the Terrell’s approximation for the distributions of fission neutron numbers for $^{235}$U (the second contributor for spontaneous fissions in the RCF reactor). Noticeably, the average multiplicity $\bar\nu$ are not coming from values as listed in Ensslin~\cite{Terrell, Ensslin} but are set to match the values provided by the ENDF/B-VII.1 nuclear data library~\cite{endfb71}. All of the other nuclear data used in this work were also extracted from ENDF/B-VII.1.
The simulation campaigns were conducted at the CCRT~\cite{ccrt} computing center, using Intel\textsuperscript{\textregistered} Xeon\textsuperscript{\textregistered} E5-2680 cores. A typical Monte Carlo run required $10^5$ processor.days to simulate the startup of the reactor until convergence of the neutron population.

\begin{figure}[h]
    \centering
    \includegraphics[width=16cm, height=13cm]{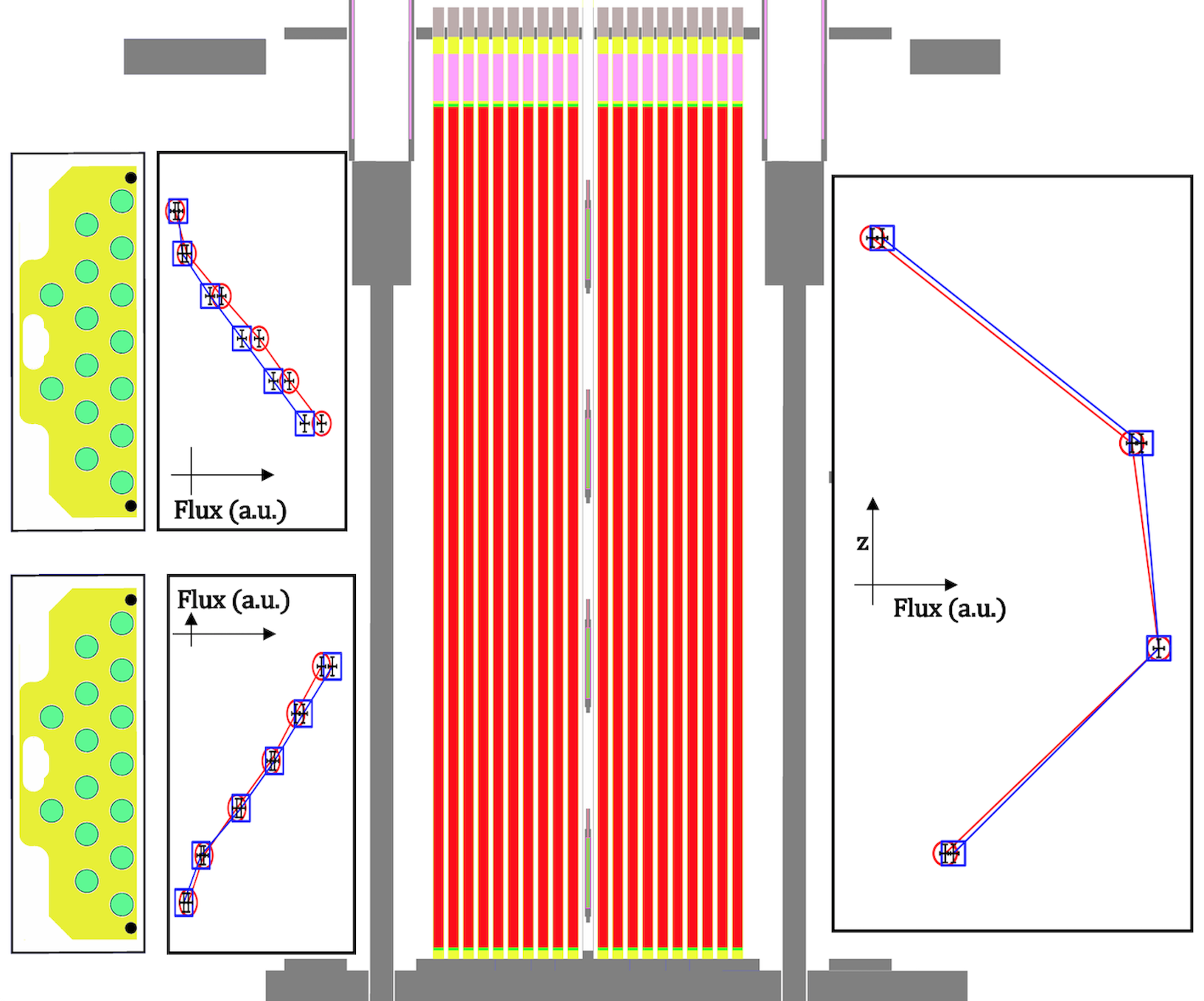}
    \caption{Comparison between the axial flux measured at the RCF (blue squares) at P=0.66 mW and the axial flux from MORET6 simulations (red circles) at P=0.79 mW for the NOMAD outer detectors (left part of the Figure) and the $^3$He inner detectors (right part of the Figure). Both axial "cosine" shapes are within 2-sigma agreement for the inner $^3$He detectors (statistic and systematic error bars are contained in the squares and circles symbols) while a systematic underestimation of the flux (by the simulation) appears on the top NOMAD detector. This can be attributed to the uncertainty on the positioning of the ex-core detectors (statistical and systematic error bars not represented).}
    \label{fig:2_All_He_counts}
\end{figure}

During the experiments and their subsequent analysis, a particular attention was devoted to the power calibration of the reactor during operation. All of the detection systems at disposal were utilized, namely the uncompensated ion chambers, the dedicated ex-core and in-core detection systems (using respectively the two NoMAD detectors and the four $^3$He tubes, all positioned along the axial dimension of the reactor core). The typical accuracy of the reconstructed power, in combination with the geometric setup of the detection system and the time performance of its electronics allowed for detailed measurements of neutron flux spatial and temporal distribution: the counting rates in the in-core and ex-core $^3$He for a 10 minute run at $P = 0.66$ mW were compared to the counting rate of a MORET6 simulation with a power value of $P = 0.79$ mW\footnote{The power value used for the simulations cannot exactly be set to the measured reactor power, as it can be only adjusted by guessing the fine tuning of the simulation parameters. Since the simulation time required to reach the asymptotic time behavior of the power can be extremely long, particularly when the core is close to critical, this procedure is rather delicate. Hence, among the set of performed simulations, we have chosen the one which had the computed power level closest to the measured power level, for the purpose of our comparisons.}. The calibrated neutron densities as computed by MORET6 are given in Figure~\ref{fig:2_All_He_counts} and show a 2-$\sigma$ agreement with the experimental data, for both in-core and ex-core detectors.

\section{A nuclear reactor "on average"}
\label{subsec:dettostoch}

So as to access average physical quantities characterizing the reactor using a simplified, yet representative, stochastic model of neutron physics, we retain the key mechanisms of diffusion, fissions on heavy nuclei, captures on absorbing materials, and spatial leakages out of the system. Intrinsic sources, like spontaneous fissions, also play a role, as shown in the following. While ($\alpha$,n) reactions contribute to the neutron background when the reactor is off, the main component of intrinsic neutron sources in the RCF core was shown to be related to spontaneous fissions on $^{238}$U (which emits $10^{-2}n/s/g$, i.e., 20 times higher than $^{235}$U). Adding a constant neutron production rate that mimics spontaneous fission to this branching Brownian model of neutron transport, the behaviour of the neutron population $n({\bf r}, t)$ at location ${\bf r}$ and time $t$ can be characterized by considering the average behavior of the probability equations, as done in Supplementary Material Sections A, B and C. In this simplified model, the ensemble-averaged neutron density $c({\bf r}, t) = \langle n({\bf r}, t) \rangle$ satisfies
\begin{equation}
\label{Master_mean2_recall}
\begin{split}
\frac{\partial}{\partial t} \, c({\bf r},t) = \bigg[ D \, \nabla^2 + \frac{\rho}{\Lambda} \bigg] c({\bf r},t)+\lambda_{SF}^v \, \overline{\nu_{SF}}
\end{split}
\end{equation}
where $D$ is the neutron diffusion coefficient, $\rho$ is the reactivity (i.e. the neutron production rate, by fissions, minus the capture rate), $\Lambda$ is the mean generation time (i.e. the average time between two generations), $\lambda_{SF}^v $ is the volumic spontaneous fission rate and $\overline{\nu_{SF}}$ is the mean number of neutrons produced per spontaneous fission. Equation~\eqref{Master_mean2_recall} has the form of a standard reaction-diffusion equation. Instead of considering the full-3D solutions of this equation, it is convenient to investigate the system behaviour along the vertical ($z$) axis, i.e., the coordinate explored by the experimental detectors in RCF: this is achieved by projecting the solution $c({\bf r},t)$ along the selected Cartesian axis, to yield $c(z,t)$. Stationary solutions can be determined assuming a negative reactivity and leakage boundary conditions at the extremities of the one-dimensional core $\pm L$
\begin{equation}
\label{sol_stat_sources}
\begin{split}
c(z)=\frac{n_\infty}{2 L} \Big(1-\frac{\cosh{(z/L^*)}}{\cosh{(L/L^*)}} \Big)
\end{split}
\end{equation}
where we have introduced the rescaled length $L^*=\sqrt{\frac{\Lambda D}{-\rho}}$ and the asymptotic neutron number $n_\infty = -\frac{\lambda_{SF} \overline{\nu_{SF}}\Lambda }{\rho}$. 
The asymptotic neutron concentration being proportional to the reactor power $P$, the reactivity is related to the ratio between the power associated to intrinsic sources $S=\lambda_{SF} \overline{\nu_{SF}}$ and the system power $P$ through $\rho=-S/P$: $\rho$ therefore ranges from $\rho = -1$, when sources are not amplified by multiplication in the core, up to $\rho = 0$, when $P$ diverges due to the amplification of the constant production of neutrons by intrinsic sources.
In the absence of sources, Eq.~\eqref{Master_mean2_recall} admits the stationary solution $c(z)=A \cos{\Big(\frac{\pi z}{2 L}\Big)}$ where $A$ is an arbitrary amplitude, provided that $\rho>0$ satisfies the condition $L=\frac{\pi}{2}\sqrt{\frac{\Lambda D}{\rho}}$.
Depending on the specific reactor power, the spatial shape of the neutron density can be driven by intrinsic sources (exponentially decaying at the boundaries, as in Eq.~\eqref{sol_stat_sources}) or by the exact balance between specific geometrical and material properties of the reactor (the cosine shapein the previous equation), referred to as the exactly critical state of the core.
In the case of the RCF reactor, above $P =1$ W, the reactor exhibits this cosine shape, while below $P =10$ mW it is driven by spontaneous fissions and the spatial shape of the neutron density can be fitted by Eq.~\eqref{sol_stat_sources}). One may be tempted be to think that this transition characterizes the transition from a deterministic regime, at high power, down to a stochastic regime, at lower power (which will be studied in the following). But we will see that both regime are closely intertwined, and cannot be differentiated considering the power solely. 
Since the previous physical quantities describe a reactor on (statistical) average, more information concerning a particular reactor can be extracted from the calculation of fluctuations and spatial correlations affecting the spatial shape of the neutron density. In the following, both experimental and numerical aspects used to measure these quantities will be discussed. 

\section{Fluctuations and neutron blinking}
\label{subsec:fluctuations}

\begin{figure}[h]
    \centering
    \includegraphics[width=17cm,height=11cm]{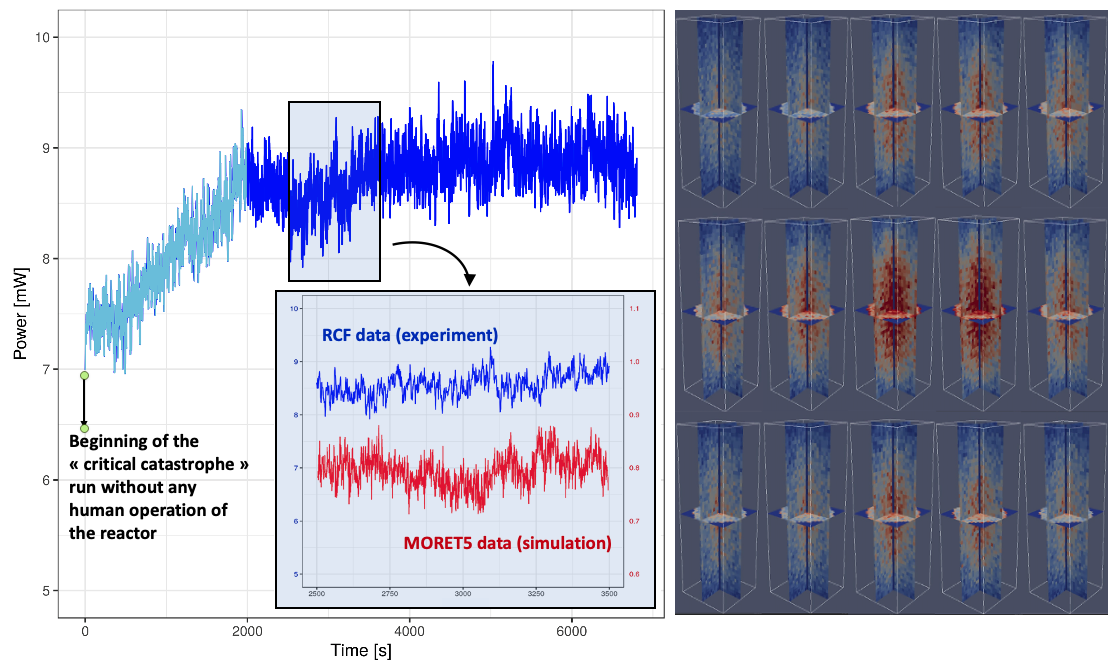}
    \caption{Left plot: "Critical catastrophe" run (in blue) at 9 mW during the RCF experiments, compared to MORET6 simulated data (in red) at 0.8 mW. The time-gate width used by the offline analysis was set to 1 ms. After a transient period (light blue) where the system reaches equilibrium (Eq.~\eqref{EqSolNis}), both signal stabilize as well as their fluctuations. While the power levels of the RCF data signal and the MORET6 simulated data are different, the stochastic neutron noises (variance-to-mean) are approximately the same, as foreseen by Eq.~\eqref{EqFluctuation}. The "critical catastrophe" foreseen by Williams~\cite{williamsBOOK} is prevented by the mechanism described Eq.~\eqref{EqSolNis}: the fluctuations are contained by the under-critical behavior of the core related to spontaneous fission sources. Right plot: Fluctuations of the power within the reactor. The fully "analog" simulation of the entire setup operating at 1.2 mW (reactor equipped with detectors and simulated with realistic statistics of neutrons) has been performed during $10^4$ processors.day on Intel\textsuperscript{\textregistered} Xeon\textsuperscript{\textregistered} E5-2680 cores. Whenever analyzed with 1 ms time gate, the core exhibits a "blinking power" behavior, as shown on this 15 ms sequence of successive snapshots. The internet link to the full video can be found here: http://eric.dumonteil.free.fr/publications/rcf.mp4.
    }
    \label{fig:3_Critical_cata}
\end{figure}

\begin{figure}[h]
    \centering
    \includegraphics[width=9cm,height=12cm]{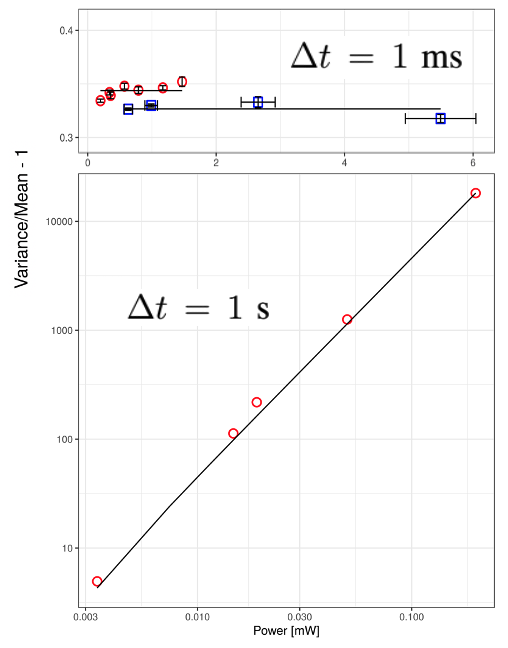}
    \caption{Variance to mean ratio versus reactor power $P$ using the inner $^3$He tubes, for various reactor power levels (simulations: red circles; RCF experiments: blue squares; theory: constant fit) for small time gate (top plot, $\Delta t=1$ ms) and for large time gate (bottom plot, $\Delta t=1$ s). In this last plot, only simulation results are presented as experimental acquisition time for such time gates were out of reach. For small time gates the noise saturates while, for large time gates, it grows unbounded following a $P^2$ law (up to the recovering of the "critical catastrophe" regime) due to vanishing intrinsic sources.}
    \label{fig:4_Critical_cata2}
\end{figure}

Indeed, provided that the reactor core is likened to an exactly critical infinite medium, the same branching Brownian model of neutron transport leads to an expression for the stochastic noise of the neutron population that can be recast in the form of the following variance-to-mean ratio (see developments exposed in Supplementary Material A)
\begin{equation}
\label{EqVtM}
\begin{split}
\frac{V_n(t)}{\langle n _ { t } \rangle}=\frac{\lambda_{F} \overline{\nu_{F}(\nu_{F}-1)}}{n_0}\ t
\end{split}
\end{equation}
where $\lambda_{F}$ is the fission rate, $\overline{\nu_{F}}$ is the average number of neutrons produced per fission, and $n_0$ is the initial number of neutrons. 
As noticed by Williams~\cite{williamsBOOK} more than 40 years ago, this variance-to-mean ($V/M$) ratio of the neutron population in the core diverges linearly with time and a critical reactor can therefore be subject to a "critical catastrophe", namely unbounded power fluctuations (eventually leading to shutdown). The main assumption is that no feedback mechanisms prevent these fluctuations from occurring. Indeed, these feedback mechanisms, such as the temperature effect at high power or operator-induced control rod movements at low power, supposedly quench the critical catastrophe (see for instance~\cite{williamsBOOK, demulatierCriticalCata, dubiCriticalCata}). 
The 2017 RCF experiments featured a dedicated run targeting the detection of this "critical catastrophe". In that aim, the RCF reactor was placed in a close-to-critical state and its power fluctuations were monitored without any human intervention to stabilize the core during time windows ranging from thirty minutes to two hours. The left plot of Figure~\ref{fig:3_Critical_cata} presents in blue the power fluctuations measured by the LP detectors of the 9 mW critical catastrophe run with a time gate width set to 1 ms during the analysis and where, after half an hour, the signal reaches its stationary 9 mW regime with a constant noise level. This measured signal was shown to be coherent with MORET6 simulations and is specifically compared to such a simulation with a power set to 0.8 mW (red curve). As seen in this figure, the variance-to-mean ratio of both power signals are close to each other and stay bounded, excluding any "critical catastrophe" driven by a stochastic drift. However, the observation of this bounded variance without any feedback was a strong indication that intrinsic sources were playing a key role in the analysis of fluctuations. Indeed, the presence of spontaneous fissions prevents the number of neutrons (and thus the power) to go down to extinction while, at the same time, requires to have a negative reactivity $\rho$ (both experimental and numerical systems are prepared so as to reach an asymptotic level). This negative reactivity also prevents the neutron population to step too far beyond its average value. The random walk of the neutron population therefore "bounces" between these virtual limits and, considering that the mean number of neutrons generated per fission and spontaneous fission are equivalent ($\overline{\nu_{SF}} \approx \overline{\nu_{F}}$), the time-asymptotic variance-to-mean ratio can be reformulated as
\begin{equation}
\label{EqFluctuationInstantaneous}
\frac { V _ { n } ( t \to \infty ) } { \langle n _ { \infty } \rangle  } = 1 + \frac { 1 } { 2 } \frac {  \: \overline { \nu _ { F } \left( \nu _ { F } - 1 \right) } } { \overline { \nu_F}  } \Big(1-\frac{1}{\rho}\Big).
\end{equation}
In this expression, the terms that add up to the Poisson noise are respectively associated to the noise of spontaneous fissions (which do not depend on reactivity) and of the fission process (inversely proportional to the reactivity $\rho=-S/P$). For a largely sub-critical reactor driven by intrinsic sources ($\rho \approx -1$), the Poisson noise is almost doubled due to the addition of a constant term characterizing the noise of the fission process itself and equal to the mean number of pairs divided by the mean number of neutrons (for spontaneous or induced fission on $^{235}$U or $^{239}$Pu this term is close to unity). It can be noted that a constant $V/M$ ratio is associated to $1/\sqrt{P}$ vanishing fluctuations $\sigma_n/{\langle n_\infty \rangle}=\sqrt{V_n}/{\langle n_\infty \rangle}$ since $\langle n_\infty \rangle \propto P$. On the contrary, when the core is close to criticality, the asymptotic variance-to-mean ratio scales as $P$ since $V _ { n }/\langle n _ { \infty } \rangle \propto \rho^{-1} \propto  P/S$, which straightforwardly leads to power-independent fluctuations $\sigma_{ n }/\langle n _ { \infty } \rangle \propto S^{-1/2}$. This is observed both in the left plot of Figure~\ref{fig:3_Critical_cata}, where the fluctuations associated to the 0.8 mW MORET6 data are reduced by a factor of three compared to the 9 mW RCF data ($\sigma_{ n }/\langle n _ { \infty } \rangle \approx 5 \%$ and $14\%$ respectively), and in the upper plot of Figure~\ref{fig:4_Critical_cata2}, where simulated and experimental data are compatible with a power-independent noise, for a time-gate width of 1 ms. However, using the set of simulated data to access the full 3D distribution of neutrons within the core led to a surprising observation: as shown in the right plot of Figure~\ref{fig:3_Critical_cata}, most of the time, the power is close to zero, and from time to time strong neutron bursts are delivered by the reactor, triggered by fission chains originating from spontaneous fission events. Observed on larger time scales (see the link to the video of the RCF MORET6 simulation), the reactor therefore presents a characteristic "blinking" behavior which signs these fluctuations driven by intrinsic sources.
Another insightful observation can be drawn from the simulated data: increasing the time-gate width $\Delta t$ from 1 ms to 1 s drastically changes the behavior of the stochastic neutron noise versus power, which follows a square law (see bottom plot of Figure~\ref{fig:4_Critical_cata2}). This can be understood by adding to the modeling of fluctuations a characteristic time scale $\alpha_d$ associated to delayed neutrons. Indeed, while prompt neutrons are emitted almost instantaneously by induced or spontaneous fissions, delayed neutrons are sometimes produced when weak forces are used by induced or spontaneous fission products (called precursors) to gain stability. Inasmuch as the beta decay occurs on much longer time scales than the emission of prompt neutrons (ranging from 1 ms up to few minutes), and even if delayed neutron represent only a small fraction $\beta$ of fission neutrons (the delayed neutron fraction $\beta$ is roughly 0.7\% for thermal fissions on $^{235}$U), both the kinetic parameters of the reactor and the time fluctuations of the neutron population are strongly affected. Integrating delayed neutron in the modeling thus requires to add a time detector (representing the real sensitive $^3$He detectors which integrate neutron capture events within time bins) so as to count the asymptotic number of detected neutrons $\langle z_\infty \rangle=\epsilon \lambda_F n_{\infty} \Delta t$ within a given time-gate $\Delta t$ ($\epsilon$ is the efficiency of the detector setup and $n_{\infty}$ is the asymptotic number of neutrons in the reactor).
Under rather mild assumptions (detailed in the Supplementary Material B), and introducing the Diven factor  $D_{\nu}=\frac{\overline{\nu_n (\nu_n -1)}}{\overline{\nu_n}^2}$ and the mean number of neutrons emitted by the decaying precursor $\overline{\nu_{m}} = \sum_{j} j \, p_{j}$, the time-asymptotic variance-to-mean ratio of the number of detected neutrons $z_\infty$ takes the form
\begin{equation} \label{FinalVtM}
     \Big\{ \frac{\text{Var}_z(t \to \infty )}{\langle z_{\infty} \rangle} \Big\}_d = 1 + \frac{\epsilon D_{\nu} (1-\rho)}{\rho^2} \left( 1 + \frac{2\, \overline{\nu_n} \, \overline{\nu_m} }{ \overline{\nu_n (\nu_n -1)} } \right)  , 
\end{equation}
Once again, these equation takes into account the stochasticity induced by spontaneous fissions via the term $(1-\rho)$. Since the average power of the reactor is directly proportional to the asymptotic number of neutrons $P \propto n_\infty \propto \rho^{-1}$, it follows that the variance-to-mean ratio follows $\text{Var}_z(\Delta t)/\langle z_{\Delta t} \rangle \propto P^2$ and, given that the average number of detected neutrons is proportional to $P$ through $\langle z_\infty \rangle=\epsilon \lambda_f n_{\infty} \Delta t$, the ratio of the standard deviation to the mean diverges when the core power is increased according to $\sigma_z(\Delta t)/\langle z_{\Delta t} \rangle \propto \sqrt{P/\Delta t}$. 
This suggests that, even though the intrinsic sources prevented the "critical catastrophe" -i.e. the unbounded growth in time of fluctuations within the neutron population-, the time scale of emission of the delayed neutrons might lead to a "delayed sub-critical catastrophe" since the system has enough time to enhance fluctuations which grow unbounded with power. This, however, can only occur for time gates width eventually not accessible to experiments. Indeed, since the numerical values of $\beta$ and $\lambda_D$ are such that $\beta / \lambda_D \approx 1$ s$^{-1}$, for close-to-critical systems we have $\alpha_d^{-1}\approx \rho^{-1} \approx P/S$ s$^{-1}$. For power reactors  during post-refueling cycle startup (operation at small $P/S$ due to low nominal power in the presence of assemblies with high burnup), one has therefore to be careful to adequately dimension the time gate over which the signal is integrated, so to ensure that the numerical protection system -which computes the neutron flux time-derivative- will not trigger unwanted scramming. Strikingly, both these formal and numerical results directly confirm that fluctuations in a reactor might grow as $\sqrt{P}$, as long as the time gate width is larger than the inverse of the delayed neutron time constant. Since this curve allows extracting numerical values such that $\sigma_z(\Delta t)/\langle z_{\Delta t} \rangle \approx \sqrt{P}$ (with $P$ in MW), at $P=1$ MW fluctuations could be of the same order of magnitude than the power itself.

\section{Spatial correlations and neutron clustering}  \label{clustering}

\begin{figure}[h]
    \centering
    \includegraphics[width=17cm, height=8cm]{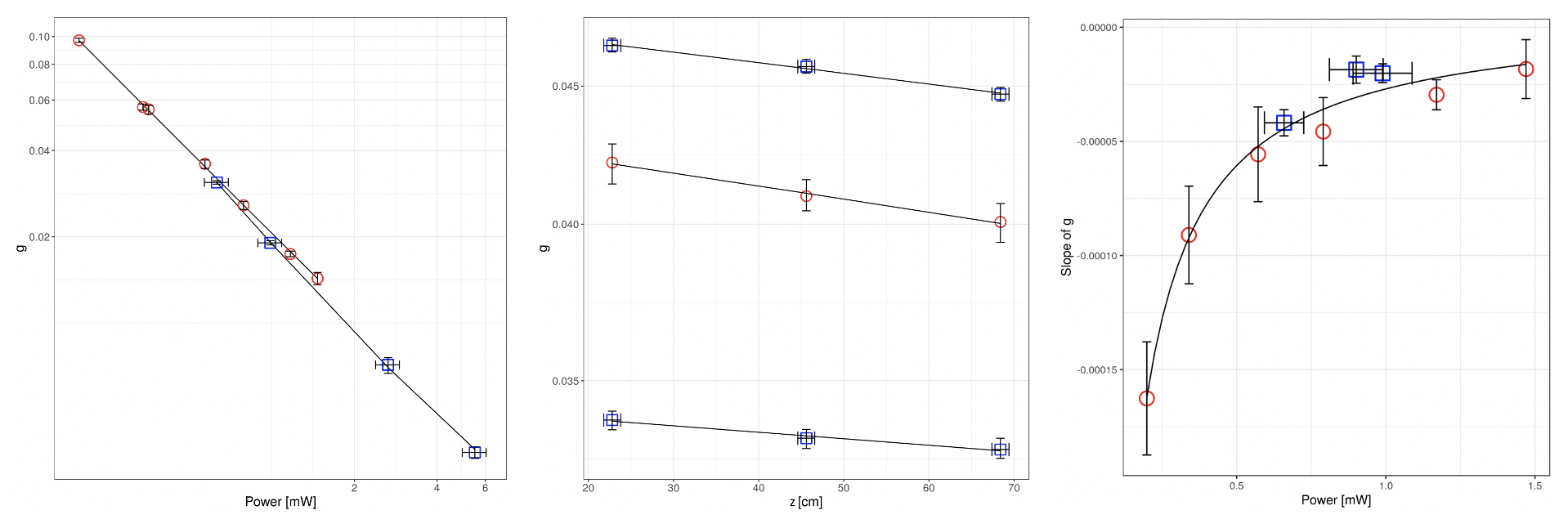}
    \caption{Left plot: spatial correlation function $g$ versus reactor power $P$ on a log-log scale using the two NoMAD detectors (simulations: red circles; RCF experiments: blue squares; theory: $1/P$ fit). The spatial correlations vanishes as $1/P$, coherently the predictions of spatial clustering models. Middle plot: spatial correlation function $g$ versus distance $z$ on a lin-lin scale using the inner $^3$He tubes, for various reactor power levels. Simulated (red circles, for $P=0.79$ mW) and experimental (blue squares, bottom curve $P=0.63$ mW, top curve: $P=0.99$ mW) data are fitted using a linear function predicted by the theory. The typical spatial linear decay of the correlation function signs a clustering effect of the neutron population in the presence of intrinsic sources. Right plot: axial gradient of the spatial correlation function $\partial_z g$ versus distance $z$ on a lin-lin scale using the inner $^3$He tubes (simulations: red circles; RCF experiments: blue squares; theory: $-1/P$ fit). The neutron clusters sizes are inversely proportional to the reactor power.}
    \label{fig:5_G_results}
\end{figure}

\begin{figure}[h]
    \centering
    \includegraphics[width=17.5cm,height=11cm]{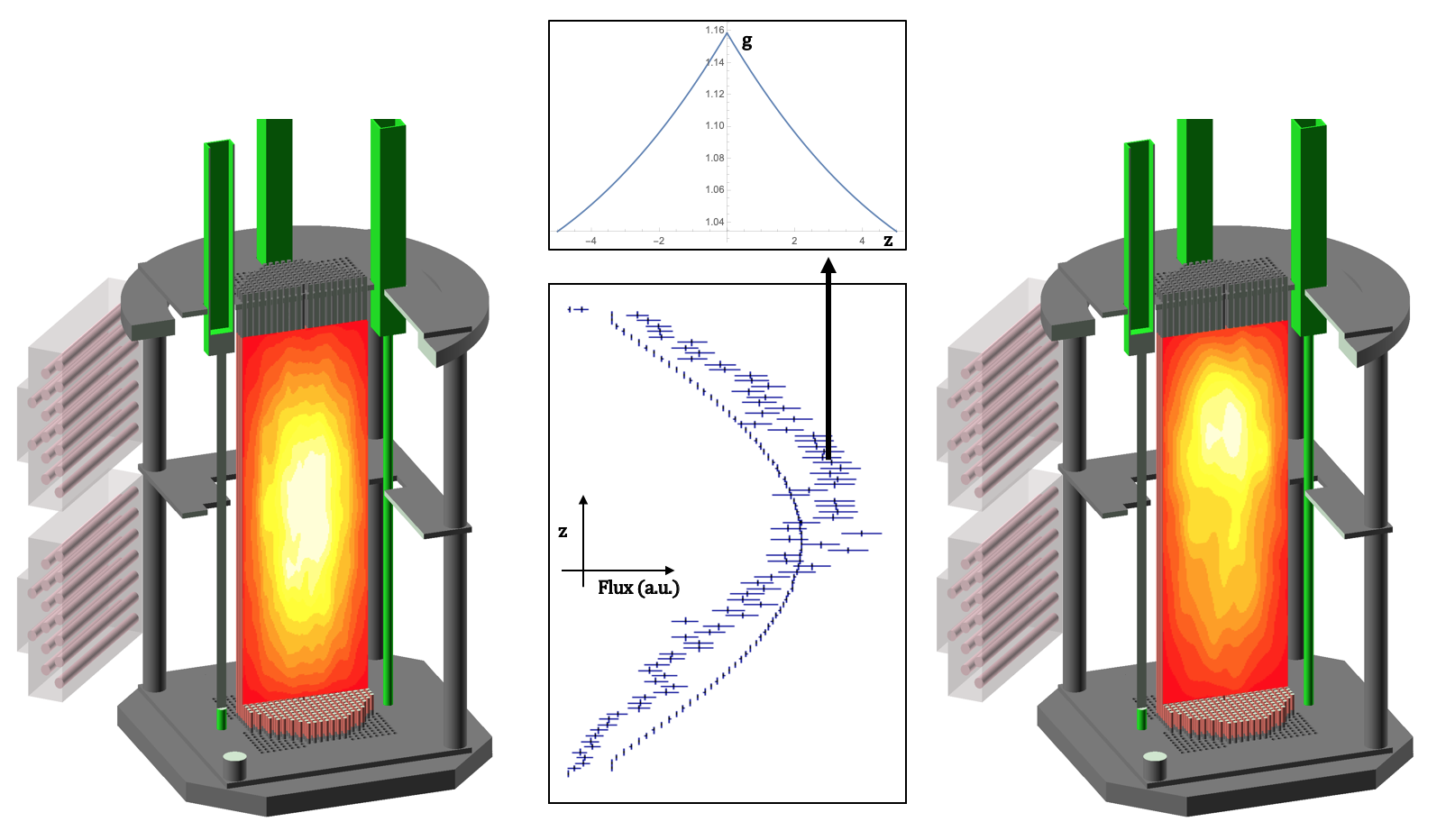}
    \caption{2D simulated power map projected onto a 3D cut of the 1.2 mW RCF experiment. Left plot: the power map is average over 100 s of acquisition time (referred to as the "averaged" power map). Right plot: the power map corresponds to an acquisition time of 1 ms (hence it will be referred to as the "snapshot" power map). The corresponding axial power profiles are superimposed in the central plot. In this last figure the  noisy profile associated to the "snapshot" view exhibits a non-Poissonian behavior typical of a clustered neutron population, that can be characterized by a peaked spatial correlation function g (central top plot).
}
    \label{fig:6_All_RCF_simu}
\end{figure}

So far, we have based our description of the power fluctuations and temporal correlations within the core mainly on the integral counting of the neutron population. For a more accurate characterization of spatial correlations, we have used the ad-hoc experimental setup described Section~\ref{sec:experiment}, allowing for simultaneous spatial measurements. The two ex-core NoMAD detectors were performing synchronized acquisition of neutron capture events in their $^3$He tubes over a wide range of reactor core powers, as their distance to the core was sufficiently large to prevent any saturation effect up to 100 mW. The experimental spatial correlation function was defined as the following two-points function
\begin{equation} \label{gNN}
    g_{N_1,N_2} = \frac{\langle N_1 N_2 \rangle-\langle N_1 \rangle \langle N_2 \rangle}{\langle N_1 \rangle \langle N_2 \rangle}
\end{equation}
where $N_1$ and $N_2$ are respectively the total number of neutrons detected in a given time bin $\Delta t$ by all $^3$He tubes of the bottom NoMAD detector (denoted $N_1$) and of the top NoMAD detector (denoted $N_2$). This function can be generalized to a 'continuous detector' model $g(x,y)$ where $x$ and $y$ are the coordinates of the 'detector' points in space. For an infinite homogeneous medium, the presence of a translation symmetry allows assuming that the spatial correlation function depends only on the distance $r$ between detectors, namely, $g(x,y)=g(r=|x-y|)$. The behavior of a neutron population evolving in such a medium has been thoroughly discussed in the literature the past years: their spatial correlation function may develop a peak for small $r$, which is related to the appearance of clustering. This phenomenon has been shown to -theoretically and numerically- persist in finite size medium with leakages, but its amplitude should be directly proportional to the reactor size (the larger the size, the longer it takes for neutrons to diffuse through the core). Other 'flavors' of clustering models have been discussed (effect of neutron energies, of delayed neutron, etc.) but ultimately all of them present two common characteristics: the amplitude of the spatial correlation function is inversely proportional to the power, while it is a decreasing function of the distance $r$. The first experimental results related to such spatial correlations are presented on the left plot of Figure~\ref{fig:5_G_results}, where the spatial correlation function $g$ defined in Eq.~\eqref{gNN} is analysed at different reactor powers. The simulation time required to numerically reproduce the different experimental runs (ranging from 0.6 mW up to 5.5 mW) were prohibitive, but it was nonetheless possible to have a partial recovering of the data between 0.6 and 1.5 mW. Both experimental RCF and numerical MORET6 spatial correlation function are in excellent relative agreement. A $1/P^\alpha$ fit of both curves shows also an excellent agreement with the law $g\propto1/P$ (simulated data: $\alpha=0.96\pm0.046$, experimental data: $\alpha=1.11\pm0.044$).
Concerning the behavior of spatial correlations as a function of the distance $r$ between detectors, the positioning of the NoMAD detectors did not allow accurate measurements at very low powers due to the strong attenuation of the neutron signal associated to their positioning, therefore an empty pin-cell equipped with four $^3$He tubes was placed directly at the center of the reactor. The bottom $^3$He detector taken as a reference, Eq.~\eqref{gNN} is used to perform 3 measures of $g$ at increasing distances (between 20 cm and 70 cm far from this reference detector) and allowed accurate measurements of $g(z)$ at very low power, below the tube saturation threshold (estimated at 5 mW). Here, $z$ refers to the projection of $r$ upon the axis of revolution of the reactor, since the $^3$He tubes detection device only allowed the detection of correlations integrated over xy-slices of the core. The central plot of Figure~\ref{fig:5_G_results} presents two experimental runs (bottom curve: $P=0.63$ mW, top curve: $P=0.99$ mW) and one simulation result at $P=0.79$ mW. The three correlation functions are ordered from smaller to higher powers and seem to exhibit a linear decrease as a function of $z$. While the decreasing behavior signs a clustering trend within the neutron population, this decreasing linear function is not coherent with other neutron clustering models predicting steeper decay, e.g. following various $\Gamma(r)$ functions \cite{dumonteilANNALS,zoiaPRE,houchPRE2}. To interpret these observations we assume that, similarly to their effect on fluctuations, intrinsic sources might also have an impact on spatial correlations (indeed these sources are distributed uniformly over the reactor and could therefore flatten the spatial correlations). For this reason, it is possible to extend the stochastic model described in Section~\ref{subsec:dettostoch} in order to include the effects of spatial correlations. The derivation presented in the Supplementary Material Section C leads to
\begin{equation}
\label{g3Dr}
g^{3D} _ { \infty } ( r ) =  \frac { \lambda _ { F } \overline { \nu _ { F } \left( \nu _ { F } - 1 \right) } } { 8 \pi D } \: \frac{ \exp{\Big( -\sqrt{\frac{\lambda _ { SF }^v \: \overline { \nu _ { SF }  }}{D \: c_\infty}} \: r \Big)}}{c_\infty \: r}
\end{equation}
where $D$ is the neutron diffusion coefficient (in cm$^2$-s$^{-1}$), and $c_\infty$ is the neutron concentration (cm$^{-3}$). The power of the reactor being linearly proportional to $c_\infty$, the spatial correlations within the core are decreasing by increasing the core power. If one uses the central pincell equipped with $^3$He detectors to measure correlations, since this pincell is placed along the $z$-axis (the axis of revolution of the core), the $g(r)$ function must be projected on this axis and is therefore given by 
\begin{equation}
\label{g3Dz}
g^{3D} _ { \infty } ( z ) =  \frac { \lambda _ { F } \overline { \nu _ { F } \left( \nu _ { F } - 1 \right) } } { 8 \pi D c_\infty} \Big(2 L_{T} \sinh ^{-1}(1)-\frac{\pi}{2}z \Big).
\end{equation}
where $L_{T}$ is the typical effective transverse dimension of the reactor, assumed to be bigger than its axial dimension (while the transverse dimension of the RCF core is roughly 30 cm and its axial dimension is $L\approx100$ cm, the axial pin-cell structure of the reactor is such that $L_T>>L$ since the fuel is homogeneous along $z$ while being decoupled by the Dancoff effect along its transverse dimension, where fuel cells are separated by water, as precised in Supplementary Material Section C).
It is remarkable that the slope of the correlation function along the $z$ axis depends on the power of the reactor via $c_\infty$ 
\begin{equation}
\label{dg3D}
\frac{\partial}{\partial z} \: g^{3D} _ { \infty } =  - \frac { \lambda _ { F } \overline { \nu _ { F } \left( \nu _ { F } - 1 \right) } } { 16 D} \: \frac{1}{c_\infty}
\end{equation}
which implies that this slope, normalized by the correlation function at $z=0$, only depends on the typical transverse size of the system through
\begin{equation}
\label{dg3Dsg}
\frac{\partial}{\partial z} \: g^{3D} _ { \infty } / \big\{g^{3D} _ { \infty }\big\}|_{z=0}=-\big( 4 \pi L_T \sinh ^{-1}(1) \big)^{-1} \approx -L_T^{-1}.
\end{equation}
Both the predictions that the slope decays as the inverse of $P$ and that it is linear are confirmed in the left and central plots of Figure~\ref{fig:5_G_results} respectively. The agreement between experimental correlations and numerical correlations lies within 2-$\sigma$ and a linear fit of $g(z)$ for numerical results is presented to also reveal the excellent agreement with the prediction of the model of neutron clustering with intrinsic sources given by Eq.~\eqref{g3Dz}. Eq.~\eqref{g3Dz} also foresees that the typical size of neutron clusters should be given by the inverse of the slope $\big(\partial g/\partial z \big)^{-1}$ and hence should increase with $P$ up to reaching the size of the core, which is verified by numerical and experimental results in the right plot of Figure~\ref{fig:5_G_results}. In particular, a strong prediction of the model is that, whenever the core is more decoupled along its transverse dimension (and hence $L_T \gg L$), the size of the clusters along the axial dimension normalized by the correlations themselves should be of the order of $L_T$. Due to the heterogeneous radial structure of the reactor, this hypothesis is hard to verify and set limits to the predictive capabilities of a model based on the assumption of a homogeneous infinite medium. However, the numerical results for $P=0.79$ mW confirm the hypothesis that $L_T$ are one order of magnitude larger than $L$.
Pursuing the approach followed to describe fluctuations, this characterization of spatial correlations suggests that a snapshot of the reactor core during operation should exhibit a non-Poissonian behavior with clustered distributions of neutrons that can be spotted with naked eye ($g\approx 5$\% on average). Accessing such a snapshot can be performed numerically, using the 1.2 mW run. Figure~\ref{fig:6_All_RCF_simu} reports one of these strongly correlated distributions of neutrons through a 2D cut view of the neutron flux (right plot) that can be compared to its time average equivalent (left plot). The center plot shows the projection of both 3D neutron flux maps on the $z$-axis. This "instantaneous" (i.e. snapshot) view is presented together with its computed (non normalized) spatial correlation function $g$ (central top plot).

\section{Conclusions}

Understanding fluctuations and spatial correlations is essential for the safe operation of nuclear reactors, especially during start-up, when stochastic effects are predominant. In the light of recent theoretical results predicting the existence of strong spatial correlations in reactors under some circumstances -the so-called neutron clustering effect-, a dedicated program led by LANL, IRSN and CEA has been developed, encompassing theoretical and numerical investigations to support experimental observations made in 2017 at the RPI Reactor Critical Facility. These observations reveal a neutron clustering phenomenon characterized by a linear decay of the spatial correlation function, and show that the reactor power is delivered through neutron "burst" from which originates a typical blinking of the nuclear core. High-fidelity simulations of the stochastic neutron transport in the reactor (taking into account the sampling of neutron fission events over realistic multiplicity and energy distributions) agree very well with both observations, and allow unravelling these clustering and blinking behavior of the core from the time-averages of the neutron population. A simplified stochastic model has been derived in order to interpret these phenomena, relying on the description of spontaneous fissions in addition to neutron-induced fissions. These spontaneous fission smooth, without completely suppressing, the tendency of neutrons to cluster -hence the linear decay of the two points function-, and prevent the neutron population from undergoing too large fluctuations, which would cause involuntary reactor shut-down (the "critical catastrophe"). However, it was shown that the stochastic noise of the reactor not only could be strongly affected by the delayed neutron fraction (and hence by burnup effects) but also might still persist in the deterministic regime, if the time gate widths used to detect the neutrons are too large (via a power driven sub-critical catastrophe), and could therefore partly explain low-frequency sudden variations of the neutron noise threshold in power reactors \cite{Seidl, Torres, Dokhane}. Concerning spatial correlations, the clustering effect has been shown to be enhanced by the decoupling of the reactor core. A dedicated experimental program to quantify the scaling of spatial correlations with the reactor power and the reactor dominance ratio could offer perspectives to understand long-standing issues related to reactor power tilts (quadrant asymmetries of the power of unknown origin). For instance, radial power tilts in large reactor seem to closely follow the behavior of spatial correlations as they appear to be positively correlated to the reactor size (or decoupling) and to decrease with the reactor power \cite{Sargeni}: this phenomenology could be compatible with a clustering triggered by local reactivity perturbations and enhanced by neutron population control using control rods \cite{demulatierJSTAT, DumonteilTW}. Finally, the fact that both fluctuations and spatial correlations are quenched by the presence of neutron leakages intimates that small/coupled reactors should present tempered neutron noise characteristics.

\begin{addendum}
 \item[Acknowledgement]
 
This work was supported in part by the DOE Nuclear Criticality Safety Program, funded and managed by the National Nuclear Security Administration for the Department of Energy, as well as by IRSN.
The authors would also like to thank Dr. Peter Caracappa for his help with the experiment, Dr. Yaron Danon for his support, and the Senior Reactor Operators who operated the reactor during the experiments: Glenn Winters, Dr. Jason Thompson, Emily Frantz, and Alexander Roaldsand. 

\item[Author Contributions] All authors contributed equally to this work.

\item[Competing Interests] The authors declare that they have no competing financial interests.

\item[Data availability]
The data that support the findings of this study are available on request from the corresponding author E.D. The data are not publicly available due to them being generated using export-controlled codes.

\item[Code availability]
MCNP6 code was used to calibrate detectors and during the design phase of the experiment. MORET5 and MORET6 codes were used to design the experiment and to analyse the data. All codes are under export control regulations but are available upon request from https://rsicc.ornl.gov and from the OECD Data Bank service https://www.oecd-nea.org/databank/. MORET6 is the next major release of MORET codes serie and is available upon request at IRSN.

\item[Correspondence] Correspondence and requests for materials
should be addressed to E. Dumonteil~(email: eric.dumonteil@cea.fr).

\newpage

\section*{}


\newpage
\section*{Supplementary material A : Instantaneous fluctuations with and without intrinsic sources} \label{InstantaneousFluctuations}
Following closely the approach detailed by Williams~\cite{williamsBOOK}, we derive in this supplementary material section a simple model to describe the stochastic behaviour of the neutron population $n(t)$ evolving freely in an idealized nuclear reactor, modelled as an homogeneous fissile medium of infinite extent. Upon random collisions on the heavy nuclei of this medium, the neutrons undergo radiative capture events at rate $\lambda_C$ (number of captures occurring in the system per second and per neutron) and induced fission events at rate $\lambda_{F}$. The induced fissions lead to the production of $i$ daughter neutrons with a probability distribution $p_i$. On top of these induced fission events, spontaneous fissions can also occur, at rate $\lambda_{SF}$, but their rate does not depend on the neutron population itself. Similarly to induced fission events they are characterized by their generating function: the probability to produce $i$ neutrons whenever a spontaneous fission occurs is noted $q_i$. 
The stochastic dynamics of the neutron population is derived by relating the probability $P(n,t+dt)$ to observe $n$ neutrons at time $t+ \Delta t$ to the probability $P(n,t)$ to observe $n$ neutrons at time $t$ 
\begin{equation}
\begin{split}
P(n,t+\Delta t) & = \lambda_{F} \sum_{i=1}^{+\infty} p_i (n+1-i) P(n+1-i,t)\Delta t + \lambda_{SF} \sum_{i=1}^{+\infty} q_iP(n+1-i,t)\Delta t \\
& + \lambda_C (n+1) P(n+1,t) \Delta t + P(n,t)\bigg( 1-\lambda_{F} n \Delta t-\lambda_{SF} \Delta t  -\lambda_C n \Delta t \bigg)
\end{split}
\end{equation}
and by rewriting this expression making the time derivative explicit, we obtain,
\begin{equation}
\label{EqP}
\begin{split}
\partial_t P(n ,t) & = \lambda_{F} \sum_{i=1}^{+\infty} p_i (n+1-i) P(n+1-i ,t) + \lambda_{SF} \sum_{i=1}^{+\infty} q_i P(n+1-i ,t) \\
 & + \lambda_{c}(n+1)P(n+1 ,t) -\lambda_{F} n P(n ,t)-\lambda_{SF} P(n,t)-\lambda_{c} n P(n,t).
\end{split}
\end{equation}

The first order moment of the probability distribution is extracted by multiplying the probability distribution by $n$ and summing $n$ over all possible positive integer values which leads to the following relation for the mean number of neutrons in the system
\begin{equation}
\label{EqN}
\begin{split}
\partial_t \big \langle n \big \rangle & =  \frac{\rho}{\Lambda} \big \langle n \big \rangle+\lambda_{SF}\overline{\nu_{SF}} ,
\end{split}
\end{equation}
where we have set $\rho/\Lambda=\big[\lambda_{F}  (\overline{\nu}-1)-\lambda_{c}\big]$, and where $\rho$ is the reactivity, $\Lambda$ is the prompt neutron generation time and $\overline{\nu} = \sum_i i\ p_i$ and $\overline{\nu_{SF}}=\sum_i i\ q_i$ are respectively the mean number of neutrons emitted per induced or spontaneous fission. 
Similarly, the evolution of the variance $V_n = \langle n^2 \rangle - \langle n \rangle^2 $ of the neutron population is obtained by retrieving the second order moments of $P(n,t)$. Through multiplication of both sides of Eq.~\eqref{EqP} by $ \sum n^2$, and after using Eq.~\eqref{EqN}, we obtain a differential equation for the variance of the number of neutrons at time $t$
\begin{equation}
\label{EqV}
\begin{split}
\partial_t  V_n  & =  2 \frac{\rho}{\Lambda}  V_n +\big(\lambda_{F}\overline{\nu(\nu-1)}-\frac{\rho}{\Lambda} \big)\big \langle n \big \rangle +\lambda_{SF}\big(\overline{\nu_{SF}(\nu_{SF}-1)}+\overline{\nu_{SF}}\big)
\end{split}
\end{equation}
where $\big \langle n(t) \big \rangle$ is given by Eq.~\eqref{EqN}, where $\overline{\nu(\nu-1)}=\sum_i i(i-1)\ p_i$ is the mean number of pairs emitted per fission, and where $\overline{\nu_{SF}(\nu_{SF}-1)}=\sum_i i(i-1)\ q_i$ is the mean number of pairs emitted per spontaneous fission.
If we neglect the intrinsic sources by setting $\lambda_{SF}=0$, the integration of Eq.~\eqref{EqN} lead straightforwardly to a $\rho$ driven geometric behavior of the average number of neutrons in the system $\big \langle n(t) \big \rangle=n_0 e^{\; \rho \; t /\Lambda}$.  
As the most important feature in the design of nuclear reactors is their ability to self-stabilize the power (and hence the instantaneous number of neutrons) at which they are operated, the reactor core can be seen as an object being designed to adjust its geometry (using control rods) and its composition (boron concentration) accordingly. At nominal powers, different feedback effects (like the Doppler effect or positive void coefficients) are used to ensure this crucial safety function. At low power however, these phenomena are not effective and the average neutron population in the core can only be adjusted thanks to a fine tuning of the reactivity $\rho$ through the use of control rods or adjustment of the boron concentration fo. Since intrinsic sources are neglected, it means that the branching process is exactly critical ($\lambda_C=\lambda_{F} (\overline{\nu} -1)$ and $\rho \sim 0$) and therefore  $\big \langle n(t) \big \rangle$ stays constant, while the variance-to-mean ratio diverges linearly with time 
\begin{equation}
\label{EqVtM}
\begin{split}
\frac{V_n(t)}{\big \langle n(t) \big \rangle}=\frac{\lambda_{F} \overline{\nu_{F}(\nu_{F}-1)}}{n_0}\ t
\end{split}
\end{equation}
(provided that all initial configurations start with the same initial number of neutrons $n_0$). The fluctuations of the neutron population are therefore becoming of the same order of magnitude than the mean number of neutron at typical time being given by $ t_{cc} \sim \frac{n_0}{\lambda_{F} \overline{\nu_{F}(\nu_{F}-1)}}$. This almost-sure extinction of the neutron population at $t_{cc}$ was theorized by M.M.R.~Williams in the 70s \cite{williamsBOOK} and is called the "critical catastrophe". It is indeed striking to realize that while the statistically averaged number of neutrons stays constant, almost each individual "reactor" with $n_0$ initial neutrons sees its population disappear. This phenomenon was a large part of the motivation to perform an experimental program at the RCF, that will be reported in the experimental section.
Whenever spontaneous fissions cannot be neglected, a careful inspection of Eq.~\eqref{EqN} indicates that $\rho$ needs to be strictly negative, so as to ensure that the power of the nuclear reactor will stay constant (an exactly critical process could not prevent the power to linearly diverge in time under the linear additive production of neutrons coming from intrinsic sources). The full time-dependent dynamic of the neutron population can be therefore be recast under this form
\begin{equation}
\label{EqSolNis}
\begin{split}
\big \langle n (t) \big \rangle & =  n_0 e^{\rho t/ \Lambda}+ n_\infty(1-e^{\rho t/\Lambda}),
\end{split}
\end{equation}
where we have introduced the time asymptotic neutron number $n_\infty=-\frac{\lambda_{SF} \overline{\nu_{SF}}\Lambda }{\rho}$.
Starting at a given value $n_0$, the neutron population undergoes a transient and, when $t\gg \Lambda/|\rho|$, it reaches its asymptotic value $n_\infty=\frac{\lambda_{SF}\overline{\nu_{SF}}\Lambda }{|\rho|}$.
At this stage it is important to note that in reality a small fraction of induced fissions lead to delayed neutrons emission, and that these delayed neutron can be shown to have a strong impact on the time behavior. While these events will not be considered in the following, as we will mostly concentrate on asymptotic time results, it can however be noted that replacing the prompt neutron generation time $\Lambda$ by an effective neutron generation time $\Lambda_{eff}$ allows to approximate with an excellent accuracy the neutron population dynamics \cite{bellBOOK}.
At large times, keeping in mind that $\rho<0$, the integration of Eq.~\eqref{EqV} can be written using the expression of $n_\infty$ and allows to retrieve an expression of the variance-to-mean ratio that takes into account the stochasticity of the intrinsic fission source
\begin{equation}
\label{EqFluctuation}
\frac { V _ { n } ( \infty ) } { n _ { \infty } } = 1 + \frac { 1 } { 2 } \frac { \lambda _ { F } \: \overline { \nu _ { F } \left( \nu _ { F } - 1 \right) }\Lambda } { \left|\rho \right| } + \frac { 1 } { 2 } \frac { \: \overline { \nu _ { SF } \left( \nu _ { SF } - 1 \right) } } { \overline {\nu _ { SF } } }
\end{equation}
This equation generalizes the celebrated formula for the variance-to-mean ratio caused by fission alone \cite{williamsBOOK} to a nuclear reactor with stochastic spontaneous fission. It is the sum of three terms: the first one comes from the Poissonian noise associated to the randomness of fissions, the second one indicates that whenever the reactor is close to criticality, the branching process can induce an arbitrarily large variance-to-mean ratio, the last term is associated to spontaneous fission sources. For most spontaneously fissile isotopes its value is between 0.5 and 1.
Compared to the case without intrinsic sources, the fluctuations of the neutron population exhibit a qualitative behavior that is radically different since they are bounded : indeed, the fact that intrinsic sources impose a sub-critical behavior of the core at constant power prevent the population to grow too high, while the fact that intrinsic sources constantly feed the core in neutrons prevent the neutron population to drop too low.

\section*{Supplementary material B: Fluctuations with intrinsic sources and delayed neutrons} \label{TemporalCorrelations}
To gain a better understanding of the temporal structure of the stochastic fluctuations in the neutron population, we need to elaborate on the model that was previously built to describe the neutron population evolving in an infinite reactor. This model must indeed acknowledge the central role that is played by delayed neutron and its associated precursor population. In this context, the underlying stochastic process must take into account for both the total number of neutrons $n(t)$ and precursors $m(t)$, behaving like random variables with a joint probability function that, like before, can be put into equation from the basic underlying phenomena affecting any one of those populations, namely: 
\begin{itemize}
    \item ~neutron capture, inducing transitions of the type $\{n,m\} \rightarrow \{n-1,m\}$. We will consider in what follows that this type of event happens in the system at a constant rate of $\lambda_C$ capture events per second and per neutron. 
    \item ~neutron induced fission, with transitions $\{n,m\} \rightarrow \{ n -1 + i , m + j\}$. Here $i,j$ denote respectively the numbers of neutrons and precursors emitted by one fission. Those numbers are random variables with a joint probability $p_{i,j}$. Fission occur in the system at a constant rate of $\lambda_F$ fission per second and per neutron.
    \item ~precursor decay which we will model as a transition $\{n,m\} \rightarrow \{n+1,m-1\}$ occurring at a constant rate of $\lambda_D$ decay per second and per precursor.  
    \item ~spontaneous fission, inducing transitions of the type $\{n,m\} \rightarrow \{ n + k , m + l\}$. Like fission, we denote $k$ and $l$ as the numbers of neutrons and precursors emitted by one spontaneous fission. Those numbers are random variables and we will denote by $q_{i,j}$ their joint probability distribution. Spontaneous fission occur in the system at a constant rate of $\lambda_{SF}$ spontaneous fission per second.
\end{itemize}
It is now important to note that if one wants to compare the predictions of the model with an experiment,  one has to take into account that neither the instantaneous neutron nor precursor populations are directly measurable quantities. This is because neutron detectors do not count individual neutrons but proceed by integrating a number of counts over a certain time interval (hereafter referred to as a \textit{time gate}). To account for this behaviour, it is customary \cite{williamsBOOK} to provide the model with yet another random variable $z$ defined as the number of neutrons registered by an idealized detector in the time interval $[0,t]$. With this definition, we must add to the pre-existing list of phenomena the following event :
\begin{itemize}
\item ~neutron detection, inducing a transition $\{n,m,z\} \rightarrow \{n-1,m,z+1\}$. This event occurs at a constant rate of $\epsilon \, \lambda_F$ events per second and per neutron. 
\end{itemize}
From the above considerations, and provided a time interval $\Delta t$ sufficiently small so that only one of the listed event can occur, we can write an equation for the joint probability $P(n,m,z,t + \Delta t)$ of finding exactly $n$ neutrons, $m$ precursors and $z$ detected neutrons in the system at time $t + \Delta t$ given the probability was $P(n,m,z,t)$ at time $t$ :

\begin{equation}
    \begin{split}
    P(n,m,z,t+\Delta  t) \,  & = \, \lambda_C \Delta t (n +1) P(n+1 , m, z, t)  \, +  \, \epsilon \Delta t (n +1) P(n+1,m,z-1,t) \\
    & + \, \lambda_D  \Delta t (m+1) P(n-1,m+1,z,t)) \\
    & + \, \lambda_F \Delta t \sum_{i,j} p_{i,j} (n + 1 - i) P(n+1-i , m-j , z,t)  \\
    & + \, \lambda_{SF} \Delta t \sum_{k,l} q_{k,l}  P(n-k , m-l , z,t) \\
    & + \, (1 - \lambda_C \Delta t n - \lambda_D \Delta t m - \lambda_F \Delta t n - \epsilon \lambda_F  \Delta t n  - \lambda_{SF} \Delta t)   P(n,m,z,t).
    \end{split}
\end{equation}
Rearranging the terms, we find that $P(n,m,z,t)$ obeys a differential equation,
\begin{equation} \label{eqP}
    \begin{split}
        \frac{d}{d t} P(n,m,z,t) & =  \lambda_C  (n +1) P(n+1 , m, z, t)  \, +  \, \epsilon \lambda_F (n +1) P(n+1,m,z-1,t) \\
        & + \, \lambda_D  (m+1) P(n-1,m+1,z,t)) \\
        & + \, \lambda_F \sum_{i,j} p_{i,j} (n + 1 - i) P(n+1-i , m-j , z,t)  \\
        & + \, \lambda_{SF} \sum_{k,l} q_{k,l}  P(n-k , m-l , z,t) \\
        & - \, (\lambda_C n + \lambda_F n \, + \epsilon \lambda_F n + \lambda_D m + \lambda_{SF}) P(n,m,z,t).
    \end{split}
\end{equation}
\newline
Eq.~\eqref{eqP} is the \textit{(forward) master equation} from which all of the moments of the probability distribution can be extracted. Considering for instance the mean number of neutrons in the system at time $t$, we can inject the definition  $\langle n(t) \rangle = \sum_{n,c,z} n P(n,c,z,t)$ into Eq.~\eqref{eqP} and immediately get,
\begin{equation} \label{eqMeanN}
    \frac{d}{d t} \langle n(t) \rangle = \left[ \lambda_F (\overline{\nu_n}-1) - \lambda_C - \epsilon \lambda_F \right] \langle n(t) \rangle + \lambda_D \langle m(t) \rangle + \lambda_{SF} \overline{\nu_{SF,n}},
\end{equation}
where we have defined $\overline{\nu_n} = \sum_{i,j} i \, p_{i,j}$ and $\overline{\nu_{SF,n}} = \sum_{k,l} k \, q_{k,l}$, the mean number of neutron produced per fission and spontaneous fission respectively. Along the same line, we can obtain an equation for the mean precursor population,
\begin{equation} \label{eqMeanC}
    \frac{d}{d t} \langle m(t) \rangle = \lambda_F \overline{\nu_c} \langle n(t) \rangle - \lambda_D \langle m(t) \rangle + \lambda_{SF} \overline{\nu_{SF,m}},
\end{equation}
with $\overline{\nu_m} = \sum_{i,j} j \, p_{i,j}$ and $\overline{\nu_{SF,m}} = \sum_{k,l} l \, q_{k,l}$ being respectively the mean number of precursors produced per fission and spontaneous fission. \\
We obtain finally the mean number of counts in the detector in the interval $ [0, t ] $, 
\begin{equation} \label{eqMeanZ}
    \frac{d}{d t} \langle z(t) \rangle = \epsilon \lambda_F \langle n(t) \rangle .
\end{equation}
It is worth noting that if we define the mean generation time $\Lambda^{-1} = \lambda_F ( \overline{\nu_n} +   \overline{\nu_m} )$, the proportion of delayed neutrons $\beta =   \overline{\nu_m}/ ( \overline{\nu_n} + \overline{\nu_m} )$ and the total reactivity $\rho/\Lambda = \lambda_F (\overline{\nu_n} + \overline{\nu_m} - 1) - \lambda_C - \epsilon \lambda_F$, then Eq.~\eqref{eqMeanN} and Eq.~\eqref{eqMeanC} reduce to the celebrated point kinetic equations with sources of neutrons and precursors,
\begin{equation} \label{KineticN}
      \frac{d}{d t} \langle n(t) \rangle  = \frac{\rho - \beta}{\Lambda} \langle n(t) \rangle + \lambda_D \langle m(t) \rangle + \lambda_{SF} \overline{\nu_{SF,n}},
\end{equation}
\begin{equation} \label{KineticC}
    \frac{d}{d t} \langle m(t) \rangle  = \frac{\beta}{\Lambda} \langle n(t) \rangle - \lambda_D \langle m(t) \rangle + \lambda_{SF} \overline{\nu_{SF,m}} \, .
\end{equation}
\newline
Solutions to Eq.~\eqref{KineticN} and Eq.~\eqref{KineticC} are exponentials of the type $\exp(\alpha_{p,d} t)$, with prompt and delayed proper modes $\alpha_{p,d}$ which are found to be the roots of the in-hour equation,
\begin{align}
 \alpha_p & = \frac{\beta - \rho}{\Lambda}, \label{ap}\\
 \alpha_d & = \frac{\lambda_D \rho}{\beta - \rho}. \label{ad}
\end{align}
Because we will be interested in situations were the reactor is operating under stable power conditions, we must seek solutions to Eq.~\eqref{KineticN} and Eq.~\eqref{KineticC} where the mean neutron and precursor populations are stationary, i.e. do not evolve in time. Such a situation can only happen if the total reactivity $\rho$ is negative. A situation of negative total reactivity and constant neutron population cannot be maintained without a delicate balance involving the constant source of neutrons and precursors that constitutes the spontaneous fission terms. This balance is achieved when the populations reach their asymptotic values $n_{\infty} = \langle n( + \infty) \rangle$ and $m_{\infty} = \langle n( + \infty) \rangle$. Those asymptotics can be deduced by solving the system Eq.~\eqref{KineticN}-\ref{KineticC} when derivatives are set to zero:
\begin{equation}
    n_{\infty} =  \frac{\lambda_{SF} (\overline{\nu_{SF,n}} + \overline{\nu_{SF,m}})\Lambda}{|\rho|} ,
\end{equation}

\begin{equation}
    m_{\infty} =  \frac{\lambda_{SF} (\overline{\nu_{SF,n}} + \overline{\nu_{SF,m}}) \, \beta}{\lambda_D \, |\rho|} + \frac{\lambda_{SF} \overline{\nu_{SF,m}}}{\lambda_D}.
\end{equation}
If the detector starts counting when stationarity is achieved (i.e. with $\langle z(0) \rangle = 0$), we arrive at a particularly simple expression for $\langle z(t) \rangle$ :
\begin{equation}
     \langle z(t) \rangle =  \epsilon \lambda_F \, n_{\infty} \, t \, .
\end{equation}
\newline
Equations for the second moments of the probability distribution can also be derived from Eq.~\eqref{eqP}, following the same line of reasoning than we did for the mean equations. As earlier, the subcritical configuration of the reactor allow us to seek stationary solutions for the variance of the neutron and precursor populations. With a little of algebra, one can show that those stationary values are solutions to the system of equations

\begin{equation}
     2 \frac{\rho - \beta}{\Lambda} \, \mu_{nn} + 2 \lambda_D \mu_{nm} + \lambda_F \, \overline{\nu_n (\nu_n -1)} \, n_{\infty} + \lambda_{SF} \, \overline{\nu_{SF,n} (\nu_{SF,n} -1)}  = 0 \, ,
\end{equation}
\begin{equation}
    -2 \lambda_D \mu_{mm} + 2 \frac{\beta}{\Lambda} \mu_{nm} + \lambda_F \, \overline{\nu_m (\nu_m -1)} \, n_{\infty} + \lambda_{SF} \, \overline{\nu_{SF,n} (\nu_{SF,n} -1)} = 0 
\end{equation}
\begin{equation}
    \left(\frac{\rho - \beta}{\Lambda} - \lambda_D \right) \mu_{nm} + \frac{\beta}{\Lambda} \mu_{nn} + \lambda_D \mu_{mm} + \lambda_F \overline{\nu_{nm}} \, n_{\infty} + \lambda_{SF} \, \overline{\nu_{SF,nm} } = 0
\end{equation}
where we defined the centralised variances $\mu_{XX} = \langle X^2 \rangle - \langle X \rangle^2 - \langle X \rangle$, covariances $\mu_{XY} = \langle X^2 \rangle - \langle X \rangle \langle Y \rangle$ and the cross moments of the distribution of neutron and precursor emitted by fission $\overline{\nu_{nm}} =  \sum_{i,j} i j \, p_{i,j}$ and spontaneous fission $\overline{ \nu_{SF , nc} }  =  \sum_{k,l} k l \, p_{k,l} $. Using the stationary solutions that we obtained for the neutron and precursor populations, we can solve the sytem of differential equations that one gets for the moments involving $z$,
\begin{align} 
    \frac{d}{dt} \mu_{zz}(t) & = 2 \, \epsilon \lambda_F \, \mu_{nz}(t) , \label{eqVarZ} \\[1em]
    \frac{d}{dt} \mu_{nz}(t) & = \frac{\rho - \beta}{\Lambda} \, \mu_{nz}(t) \, + \, \lambda_D \mu_{mz}(t) \, + \, \epsilon \lambda_F \mu_{nn} , \label{eqVnz} \\[1em]
    \frac{d}{dt} \mu_{mz}(t) & = \frac{\beta}{\Lambda} \mu_{nz}(t)  \, - \, \lambda_D \mu_{mz}(t) \, + \, \epsilon \lambda_F \mu_{nm} \label{eqVCZ}.
\end{align}
The system of coupled differential equations (Eq.~\eqref{eqVarZ} - \ref{eqVCZ}) can be solved by resorting to Laplace transformations \cite{PazsitBOOK}. After a bit of tedious but rather straightforward algebra, we arrive at the celebrated variance-to-mean ratio or \textit{Feynman alpha} function
\begin{equation} \label{FeynmanAlpha}
     \frac{\text{Var}(z)}{\langle z \rangle} = \frac{\mu_{zz}(t)}{\langle z(t) \rangle} + 1 = 1 + Y_{p} \left( 1 - \frac{1 - e^{-\alpha_p t}}{\alpha_p \, t} \right)  \, + \, Y_{d} \left( 1 - \frac{1 - e^{-\alpha_d t}}{\alpha_d \, t} \right) , 
\end{equation}
where $\alpha_{p,d}$ are given by Eq.~\eqref{ap} and Eq. \ref{ad} and we have defined the two constants
\begin{equation} \label{eqYp}
    Y_p = \frac{-2 \Lambda \epsilon \lambda_F}{n_{\infty} (\alpha_p - \alpha_d) \rho} \left[ \left(\lambda_D - \alpha_p - \frac{\rho - \beta}{\Lambda}\right) ( \mu_{nn} + \mu_{nm}) + \frac{\rho}{\Lambda} \mu_{nn} \right]
\end{equation}
and
\begin{equation} \label{eqYd}
    Y_d = \frac{2 \Lambda \epsilon \lambda_F}{n_{\infty} (\alpha_p - \alpha_d) \rho} \left[ \left(\lambda_D - \alpha_d - \frac{\rho - \beta}{\Lambda}\right)( \mu_{nn} + \mu_{nm}) + \frac{\rho}{\Lambda} \mu_{nn} \right] .
\end{equation}
\newline
The complete forms of $Y_p$ and $Y_d$ is obtained upon insertion of the explicit expressions for $n$, $\mu_{nn}$ and $\mu_{nc}$ into Eq. \ref{eqYp} and \ref{eqYd}. It has been shown \cite{PazsitBOOK} that in the case where a constant source of neutrons is used in place of spontaneous fission then $Y_p$ and $Y_d$ can be written, to a very good approximation, as
\begin{equation} \label{Yp_class}
    Y_p^{\text{SF=0}} = \frac{\epsilon  D_{\nu}}{(\beta - \rho)^2} 
\end{equation}
and 
\begin{equation} \label{Yd_class}
    Y_d^{\text{SF=0}} = \frac{\epsilon  D_{\nu}}{( \beta - \rho)^2} \left[ \left(\frac{\rho - \beta}{\rho}\right)^2 \, \left( 1 + \frac{2\, \overline{\nu_{nm}}  }{ \overline{\nu_n (\nu_n -1)} } \right) -1 \right],
\end{equation}
where we have introduced the \textit{Diven factor},
\begin{equation}
    D_\nu = \frac{\overline{\nu_n (\nu_n -1)}}{\overline{\nu_n}^2}.
\end{equation}
\newline
It is possible to isolate the terms in $Y_p$ and $Y_d$ that must be added to $Y_p^{\text{SF=0}}$ and $Y_d^{\text{SF=0}}$ to take into account the stochastic nature of the neutron and precursor sources when spontaneous fission is considered. With $\Lambda \lambda_D \ll 1$ we can relate the overall Feynman-$\alpha$ $Y_p$ to the Feynman-$\alpha$ without stochasticity of spontaneous fissions $Y_p^{\textit{SF=0}}$
\begin{equation}
    Y_p = Y_p^{\textit{SF=0}} - \frac{\epsilon \lambda_F \Lambda \rho}{(\beta - \rho)^2} \, \frac{ \overline{ \nu_{SF,n} (\nu_{SF,n} -1) }  }{ \overline{ \nu_{SF,n}} + \overline{ \nu_{SF,m}} },
\end{equation}
Since it can be safely assumed that the average number of neutrons produced by beta decay of spontaneous fragment fission is small compared to the average number of neutrons produced per spontaneous fission themselves ($\overline{ \nu_{SF,n}} << \overline{ \nu_{SF,m}}$), the previous expression can be simplified to
\begin{equation}
    Y_p = Y_p^{\textit{SF=0}} - \rho \frac{\epsilon D_{\nu_{SF}} }{(\beta - \rho)^2}.
\end{equation}
where we have introduced the Diven factor for spontaneous fission $D_{\nu_{SF}} = \frac{\overline{\nu_{SF,n} (\nu_{SF,n} -1)}}{\overline{\nu_{SF,n}}^2}$.
Assuming $D_{\nu} \approx D_{\nu_{SF}}$, we arrive at a remarkably simple expression,
\begin{equation}
    Y_p = Y_p^{\textit{SF=0}}(1-\rho).
\end{equation}
\newline
Equivalently, we can write the delayed component as 
\begin{equation}
    Y_d = Y_d^{\textit{SF=0}} \, - \, \rho \frac{\epsilon D_{\nu_{SF}}}{(\beta - \rho)^2} \left[ \left( \frac{\beta - \rho}{\rho}\right)^2 \left( 1 + \frac{2 \, \overline{\nu_{SF,nm}}}{\overline{\nu_{SF,n} (\nu_{SF,n} -1)} } \right) - 1\right]
\end{equation}
and we end up with the same expression,
\begin{equation}
    Y_d = Y_d^{\textit{SF=0}}(1-\rho),
\end{equation}
\newline
As a final remark, let us finally note that by resorting to the Pluta theorem, one can show \cite{williamsBOOK} that it is possible to extract from Eq.~\eqref{FeynmanAlpha} an expression for the  \textit{auto-correlation} function $\phi_{nn}(t)$,
\begin{equation}
    \phi_{nn}(t) = \epsilon \lambda_F \, n_{\infty} \, \left\{  \delta(t) + \frac{1}{2} Y_p \alpha_p e^{- \alpha_p t} + \frac{1}{2} Y_d \alpha_d e^{- \alpha_d t} \right\} 
\end{equation}
where $\delta(t)$ is the usual delta function that accounts for the uncorrelated neutrons component.

\section*{Supplementary material C: Spatial correlations with and without intrinsic sources} \label{SpatialCorrelations}

\hfill
\newline 
We finally extend the idealized reactor model presented thus far to the study of spatial correlations in the neutron population. For simplicity, and because precursors have a much less prominent role in this case, we will restrict ourselves here to the study of the neutron population alone.
\newline \newline
The reactor model is refined by the introduction of a discrete and infinite spatial dimension in the problem. In this scenario, the neutron population is now evolving inside cells of equal length $h$ and the number of neutrons $n_k$ in each cell $k$ ($k$ is an integer) is, like before, treated as a random variable. Following the principles of the second quantization of quantum field theory, we introduce the state vector $\Vec{n}$ that is represented in its occupation number basis as well as the creation operator $a_k^+$ (that adds one neutron in cell $k$) and the annihilation operator $a_k$ (that removes one neutron in cell $k$)
  \begin{align}
    \Vec{n} &= \begin{bmatrix}
           \vdots \\
           n_{k-1} \\
           n_{k} \\
           n_{k+1} \\
           \vdots \\
         \end{bmatrix} \ \ \ \ \ \
         a_k\ \Vec{n} = \begin{bmatrix}
           \vdots \\
           n_{k-1} \\
           n_{k}-1 \\
           n_{k+1} \\
           \vdots \\
         \end{bmatrix} \ \ \ \ \ \
         a_k^+\ \Vec{n} = \begin{bmatrix}
           \vdots \\
           n_{k-1} \\
           n_{k}+1 \\
           n_{k+1} \\
           \vdots \\
         \end{bmatrix}.
  \end{align}
\newline
As in previous section, the population $\vec{n}$ of neutrons will undergo the following transitions:
\begin{itemize}
  \item ~capture, inducing transitions $\Vec{n}\rightarrow a_k \ \Vec{n}$ at rate $\lambda_C$ per second and per neutron;
  \item ~neutron induced fission (with $j$ daughters neutrons), inducing transitions $\Vec{n}\rightarrow (a_k^+)^j a_k \ \Vec{n}$ at rate $\lambda_{F} \, p_j$ per second and per neutron;
  \item ~spontaneous fission (with $j$ daughters neutrons), inducing transitions $\Vec{n}\rightarrow (a_k^+)^l \ \Vec{n}$ at rate $\lambda_{SF} \, q_j$ per second and per cell;
  \item ~we also allow for migration to occur : neutrons can travel from one cell to adjacent ones. This process induce a transition $\Vec{n}\rightarrow (a_{k\pm 1}^+)a_k \ \Vec{n}$ and occur at a rate that we define to be $\gamma$ per second and per neutron.
\end{itemize}
The master equation is once again derived by a cautious balance between the events that contribute in an increment or decrement of the neutron population in cell $k$, during a time interval $\Delta t$:
\begin{equation}
\label{EqMasterSC}
\begin{split}
\frac{\partial}{\partial t} P(\Vec{n}) & =  \sum_k \Bigg\{ -\lambda_C n_k P(\Vec{n})+ \lambda_C (n_k+1) P(a_k^+ \Vec{n}) \\ & -\lambda_{F} \sum_j p_j n_k P(\Vec{n})+ \lambda_F \sum_j p_j (n_k+1-j) P(a_k^+ (a_k)^j \Vec{n}) \\ &  -\lambda_{SF} \sum_j q_j P(\Vec{n})+ \lambda_{SF} \sum_j q_j P( (a_k)^j \Vec{n}) \\ &  -2 \gamma n_k P(\Vec{n}) + \gamma (n_{k+1}+1) P( a_{k+1}^+ a_k \Vec{n})+ \gamma (n_{k-1}+1) P(a_{k-1}^+ a_k \Vec{n}) \Bigg\}.
\end{split}
\end{equation}
In the following, the dummy index $k$ that designs cell $k$ will be replaced by a dummy index $i$ so as to not be confounded with the sum over $n_k$ neutrons in cell $k$ associated to the vector state $\Vec{n}$.
Thus, the average number of neutrons $\big \langle n_k \big \rangle $ in cell $k$ is then extracted thanks to its definition
\begin{equation}
\label{EqDefN}
\begin{split}
\big \langle n_k \big \rangle=\sum_{\Vec{n}} n_k P(\Vec{n})
\end{split}
\end{equation}
Multiplied by $n_k$ and summed over all possible $\Vec{n}$ states, the capture-related terms $\large \textcircled{\small{1}}$ in Eq.~\eqref{EqMasterSC} (first line, right hand side) can be rewritten
\begin{equation}
\label{capture}
\begin{split}
\large \textcircled{\small{1}} & = \sum_{\Vec{n}} n_k \sum_i  \Bigg\{ -\lambda_C n_i P(\Vec{n})+ \lambda_C (n_i+1) P(a_i^+ \Vec{n}) \Bigg\} \\
\large \textcircled{\small{1}} & = \lambda_C \sum_{\Vec{n}} \sum_i  \bigg\{ -n_k n_i P(\Vec{n})+  n_k (n_i+1)  P(a_i^+ \Vec{n}) \bigg\} \\
\end{split}
\end{equation}
And using the change of variable $a_i^+ \Vec{n}\rightarrow \Vec{n} $ for the state vector in the second term of the right hand side, we can notice that $n_k$ is unchanged but looses one neutron whenever $k=i$ and $n_i+1$ becomes $n_i$ 
\begin{equation}
\label{capture2}
\begin{split}
\large \textcircled{\small{1}} & = \lambda_C \sum_{\Vec{n}} \sum_i  \bigg\{ -n_k n_i +  (n_k-\delta_{k,i})n_i   \bigg\} P(\Vec{n}) \\
\large \textcircled{\small{1}} & = -\lambda_C \big \langle n_k \big \rangle
\end{split}
\end{equation}
where $\delta_{k,i}$ stands for the Kronecker delta function that is equal to $1$ only whenever $i=k$.
The fission related terms $\large \textcircled{\small{2}}$ in Eq.~\eqref{EqMasterSC} (second line) become
\begin{equation}
\label{fission}
\begin{split}
\large \textcircled{\small{2}} & = \sum_{\Vec{n}} n_k \sum_i  \Bigg\{ -\lambda_F \sum_j p_j n_i P(\Vec{n})+ \lambda_F  \sum_j p_j (n_i+1-j) P(a_i^+ (a_i)^j \Vec{n}) \Bigg\} \\
\large \textcircled{\small{2}} & =  \lambda_F \sum_{\Vec{n}} \sum_i  \sum_j \Bigg\{ - p_j n_k n_i P(\Vec{n})+ p_j n_k (n_i+1-j) P(a_i^+ (a_i)^j \Vec{n}) \Bigg\} \\
\large \textcircled{\small{2}} & =  \lambda_F \sum_{\Vec{n}} \sum_i  \sum_j \Bigg\{ - p_j n_k n_i + p_j (n_k +\delta_{k,i}(j-1)) n_i \Bigg\} P(\Vec{n}) \\
\large \textcircled{\small{2}} & =  \lambda_F  (\overline{\nu}-1) \big \langle n_k \big \rangle \\
\end{split}
\end{equation}
The spontaneous fission related terms $\large \textcircled{\small{3}}$  in Eq.~\eqref{EqMasterSC} (third line) become
\begin{equation}
\label{sfission}
\begin{split}
\large \textcircled{\small{3}} & = \sum_{\Vec{n}} n_k \sum_i  \Bigg\{ -\lambda_{SF} \sum_j q_j P(\Vec{n})+ \lambda_{SF}  \sum_j q_j P((a_i)^j \Vec{n}) \Bigg\} \\
\large \textcircled{\small{3}} & =  \lambda_{SF} \sum_{\Vec{n}} \sum_i  \sum_j \Bigg\{ - q_j n_k P(\Vec{n})+ q_j n_k P((a_i)^j \Vec{n}) \Bigg\} \\
\large \textcircled{\small{3}} & =  \lambda_{SF} \sum_{\Vec{n}} \sum_i  \sum_j \Bigg\{ - q_j n_k + q_j (n_k+\delta_{k,i}j) \Bigg\} P(\Vec{n}) \\
\large \textcircled{\small{3}} & =  \lambda_{SF}  \overline {\nu_{SF}} \\
\end{split}
\end{equation}
The migration related terms $\large \textcircled{\small{4}}$ in Eq.~\eqref{EqMasterSC} (fourth line) become
\begin{equation}
\label{migration}
\begin{split}
\large \textcircled{\small{4}} & = \sum_{\Vec{n}} n_k  \sum_i \Bigg\{ -2 \gamma n_i P(\Vec{n}) + \gamma (n_{i+1}+1) P( a_{i+1}^+ a_i \Vec{n})+ \gamma (n_{i-1}+1) P(a_{i-1}^+ a_i \Vec{n}) \Bigg\} \\
\large \textcircled{\small{4}} & = \sum_{\Vec{n}} \sum_i \Bigg\{ -2 \gamma n_k n_i P(\Vec{n}) + \gamma n_k (n_{i+1}+1) P( a_{i+1}^+ a_i \Vec{n})+ \gamma n_k (n_{i-1}+1) P(a_{i-1}^+ a_i \Vec{n}) \Bigg\} \\
\large \textcircled{\small{4}} & = \gamma \sum_{\Vec{n}} \sum_i  \Bigg\{  -2n_k n_i  + (n_k+\delta_{k,i}-\delta_{k,i+1})n_{i+1}+ (n_k+\delta_{k,i}-\delta_{k,i-1}) n_{i-1} \Bigg\} P(\Vec{n}) \\
\large \textcircled{\small{4}} & = \gamma  \big( \big \langle n_{k+1} \big \rangle + \big \langle n_{k-1} \big \rangle -2 \big \langle n_k \big \rangle \big) \\
\end{split}
\end{equation}
Using the discrete expression of the second order derivative $\Delta \big \langle n_k \big \rangle  \approx n_{i+1}+n_{i-1}-2 n_i$, $\large \textcircled{\small{4}}$ can be interpreted as the diffusion rate $\gamma \Delta \big \langle n_k \big \rangle $.
Finally, the first order moment of Eq.~\eqref{EqMasterSC} leads to
\begin{equation}
\label{Master_mean_neutron}
\begin{split}
\frac{\partial}{\partial t} \big \langle n_{k} \big \rangle =  \gamma \Delta \big \langle n_{k} \big \rangle  + (\lambda_F (\overline {\nu_{F}} -1)-\lambda_C ) \big \langle n_{k} \big \rangle+\lambda_{SF} \overline{\nu_{SF}}
\end{split}
\end{equation}
The spatial coordinate can be parameterized using $x=kh$ where we recall that $h$ is the size of a cell. Dividing Eq.~\eqref{Master_mean_neutron} by $h$ and taking the limit $h\to0$ we find
\begin{equation}
\label{Master_mean2}
\begin{split}
\frac{\partial}{\partial t} \, c(x,t) = \bigg[ D \, \frac{\partial^2}{\partial x^2} + \frac{\rho}{\Lambda} \bigg] c(x,t)+\lambda_{SF}^v \, \overline{\nu_{SF}}
\end{split}
\end{equation}
where the discrete formula for the Laplacian is replaced by the continuous second order derivative 
$$ \frac{\partial^2}{\partial x^2} = \lim_{h\to 0}  \bigg[  \big \langle n_{k+1} \big \rangle + \big \langle n_{k-1} \big \rangle -2 \big \langle n_k \big \rangle \bigg] / h^2$$
and where we have introduced the neutron concentration at position $x$ and time $t$ 
$$c(x,t) = \lim_{h\to 0} \frac{\big \langle n_{k} \big \rangle}{h},$$
the lineic spontaneous fission rate
$$ \lambda_{SF}^v=\lim_{h\to 0} \frac{\lambda_{SF}}{h},$$
and the diffusion coefficient
$$ D = \lim_{h\to 0} \gamma \, h^2$$ 
Interestingly enough, under isotropic conditions, Eq.~\eqref{Master_mean2} can trivialy be generalized to a $d$-dimensional space only be replacing the one dimensional second order derivative $\partial_x^2$ by the $d$-dimensional Laplacian operator $\Delta^d$. For an infinite medium and symmetrical initial conditions $c(x,t=0)=c_0$, the Laplacian disappears and the solution to Eq.~\eqref{Master_mean2} is found to be,
\begin{equation}
\label{Eqcxt}
\begin{split}
c(x,t)=c(t)=c_0 \, exp(\frac{\rho}{\Lambda}\, t)+c_\infty(1-e^{\rho t/\Lambda}),
\end{split}
\end{equation}
where we have introduced the time asymptotic concentration $c_\infty = -\frac{\lambda_{SF}^v \overline{\nu_{SF}}\Lambda }{\rho}$.
The mean neutron concentration does not depend on $x$ and its behavior in each point strictly follows the law without spatial dimension (with the rate of spontaneous sources being replaced by the lineic rate of the spontaneous sources).
\newline \newline
This procedure aiming at calculating the mean concentration of neutron at position $x$ and time $t$ is now closely followed once again to calculate spatial correlations, using the definition of the two-point function
\begin{equation}
\label{EqDefCor}
\begin{split}
\big \langle n_k n_{k+l} \big \rangle=\sum_{\Vec{n}} n_k n_{k+l} P(\Vec{n})
\end{split}
\end{equation}
Multiplied by $n_k n_{k+l}$ and summed over all possible $\Vec{n}$ states, the capture-related terms $\large \textcircled{\small{1}}$ in Eq.~\eqref{EqMasterSC} (first line, right hand side) can be rewritten
\begin{equation}
\label{capture2}
\begin{split}
\large \textcircled{\small{1}} & = \sum_{\Vec{n}} n_k n_{k+l} \sum_i  \Bigg\{ -\lambda_C n_i P(\Vec{n})+ \lambda_C (n_i+1) P(a_i^+ \Vec{n}) \Bigg\} \\
\large \textcircled{\small{1}} & = \lambda_C \sum_{\Vec{n}} \sum_i  \bigg\{ -n_k n_{k+l} n_i P(\Vec{n})+  n_k n_{k+l} (n_i+1)  P(a_i^+ \Vec{n}) \bigg\} \\
\end{split}
\end{equation}
And using the change of variable $a_i^+ \Vec{n}\rightarrow \Vec{n} $
\begin{equation}
\label{capture2}
\begin{split}
\large \textcircled{\small{1}} & = \lambda_C \sum_{\Vec{n}} \sum_i  \bigg\{ -n_k n_{k+l} n_i +  (n_k -\delta_{k,i})(n_{k+l}-\delta_{k+l,i})n_i   \bigg\} P(\Vec{n}) \\
\large \textcircled{\small{1}} & = \lambda_C \sum_{\Vec{n}} \sum_i  \bigg\{ -n_k \delta_{k+l,i} n_i -\delta_{k,i}n_{k+l} n_i + \delta_{k,i}\delta_{k+l,i} n_i  \bigg\} P(\Vec{n}) \\
\large \textcircled{\small{1}} & = -2\lambda_C \big \langle n_k n_{k+l}  \big \rangle +\lambda_C \big \langle n_k \big \rangle \delta_{l,0}
\end{split}
\end{equation}
where we have assumed a spatial symmetry implying that $\big \langle n_k n_{k-l}  \big \rangle=\big \langle n_k n_{k+l}  \big \rangle$.
The fission related terms $\large \textcircled{\small{2}}$ in Eq.~\eqref{EqMasterSC} (second line) become
\begin{equation}
\label{fission2}
\begin{split}
\large \textcircled{\small{2}} & = \sum_{\Vec{n}} n_k n_{k+l}  \sum_i  \Bigg\{ -\lambda_F \sum_j p_j n_i P(\Vec{n})+ \lambda_F  \sum_j p_j (n_i+1-j) P(a_i^+ (a_i)^j \Vec{n}) \Bigg\} \\
\large \textcircled{\small{2}} & =  \lambda_F \sum_{\Vec{n}} \sum_i  \sum_j \Bigg\{ - p_j n_k n_{k+l} n_i P(\Vec{n})+ p_j n_k n_{k+l} (n_i+1-j) P(a_i^+ (a_i)^j \Vec{n}) \Bigg\} \\
\large \textcircled{\small{2}} & =  \lambda_F \sum_{\Vec{n}} \sum_i  \sum_j \Bigg\{ - p_j n_k n_{k+l}  n_i + p_j (n_k +\delta_{k,i}(j-1))(n_{k+l} +\delta_{k+l,i}(j-1)) n_i \Bigg\} P(\Vec{n}) \\
\large \textcircled{\small{2}} & =  \lambda_F \sum_{\Vec{n}} \Bigg\{ (\overline{\nu -1})n_k n_{k+l}+\overline{(\nu -1)^2}n_k \delta_{l,0} + (\overline{\nu -1})n_k n_{k+l} \Bigg\} P(\Vec{n}) \\
\large \textcircled{\small{2}} & =  2\lambda_F  (\overline{\nu-1}) \big \langle n_k n_{k+l} \big \rangle +\lambda_F \big \langle n_k \big \rangle \delta_{l,0}[\overline{\nu (\nu-1)}-\overline{(\nu -1)}]
\end{split}
\end{equation}
The spontaneous fission related terms $\large \textcircled{\small{3}}$  in Eq.~\eqref{EqMasterSC} (third line) become
\begin{equation}
\label{sfission2}
\begin{split}
\large \textcircled{\small{3}} & = \sum_{\Vec{n}} n_k n_{k+l} \sum_i  \Bigg\{ -\lambda_{SF} \sum_j q_j P(\Vec{n})+ \lambda_{SF}  \sum_j q_j P((a_i)^j \Vec{n}) \Bigg\} \\
\large \textcircled{\small{3}} & =  \lambda_{SF} \sum_{\Vec{n}} \sum_i  \sum_j \Bigg\{ - q_j n_k n_{k+l} P(\Vec{n})+ q_j n_k  n_{k+l} P((a_i)^j \Vec{n}) \Bigg\} \\
\large \textcircled{\small{3}} & =  \lambda_{SF} \sum_{\Vec{n}} \sum_i  \sum_j \Bigg\{ - q_j n_k n_{k+l} + q_j (n_k+\delta_{k,i}j)(n_{k+l}+\delta_{k+l,i}j) \Bigg\} P(\Vec{n}) \\
\large \textcircled{\small{3}} & =  \lambda_{SF}  \overline {\nu_{SF}} \big \langle n_{k+l} \big \rangle+\lambda_{SF}  \overline {\nu_{SF}} \big \langle n_{k} \big \rangle +\lambda_{SF}  \overline {\nu_{SF}^2} \delta_{l,0}  \\
\end{split}
\end{equation}
The migration related terms $\large \textcircled{\small{4}}$ in Eq.~\eqref{EqMasterSC} (fourth line) become
\begin{equation}
\label{migration}
\begin{split}
\large \textcircled{\small{4}} & = \sum_{\Vec{n}} n_k n_{k+l} \sum_i -2 \gamma n_i P(\Vec{n}) + \gamma (n_{i+1}+1) P( a_{i+1}^+ a_i \Vec{n})+ \gamma (n_{i-1}+1) P(a_{i-1}^+ a_i \Vec{n}) \\
\large \textcircled{\small{4}} & = \sum_{\Vec{n}} \sum_i -2 \gamma \ n_k n_{k+l} n_i P(\Vec{n}) + \gamma n_k n_{k+l} (n_{i+1}+1) P( a_{i+1}^+ a_i \Vec{n})+ \gamma n_k n_{k+l}(n_{i-1}+1) P(a_{i-1}^+ a_i \Vec{n}) \\
\large \textcircled{\small{4}} & = \gamma \sum_{\Vec{n}} \sum_i  \Bigg\{  -2n_k n_{k+l} n_i  + (n_k+\delta_{k,i}-\delta_{k,i+1})(n_{k+l}+\delta_{k+l,i}-\delta_{k+l,i+1}) n_{i+1}+ \\ & \ \ \  \ \ \  \ \ \  \ \ \ \ \ \  \ \ \  \ \ \  \ \ \  (n_k+\delta_{k,i}-\delta_{k,i-1})(n_{k+l}+\delta_{k+l,i}-\delta_{k+l,i-1}) n_{i-1} \Bigg\} P(\Vec{n}) \\
\large \textcircled{\small{4}} & = \gamma \sum_{\Vec{n}} \Bigg\{ n_k n_{k+l+1}-2n_k n_{k+l}+n_{k+1} n_{k+l}+n_{k+1}(2 \delta_{l,0}-\delta_{l,1}-\delta_{l,-1})\\ & \ \ \  \ \ \  \ \ \  \ \ \ \ \ \  \ \ \  \ \ \  \ \ \  + n_{k}n_{k+l-1}- 2 n_{k}n_{k+l}+n_{k-1}n_{k+l}+n_{k-1}(2 \delta_{l,0}-\delta_{l,1}-\delta_{l,-1}) \Bigg\} P(\Vec{n}) \\
\large \textcircled{\small{4}} & = \gamma \sum_{\Vec{n}} \Bigg\{ n_k \Delta n_{k+l}+\Delta n_{k} n_{k+l}+\delta_{l,0}(n_{k+1}+n_{k-1}+2 n_k)\\ & \ \ \  \ \ \  \ \ \  \ \ \ \ \ \  \ \ \  \ \ \  \ \ \  -\delta_{l,1}(n_{k+1}+n_{k})-\delta_{l,-1}(n_{k-1}+n_{k})  \Bigg\} P(\Vec{n}) \\
\large \textcircled{\small{4}} & = \gamma \big \langle n_k \Delta n_{k+l} \big \rangle+\gamma \big \langle n_{k+l} \Delta n_{k} \big \rangle+\gamma \delta_{l,0}(\big \langle n_{k+1}\big \rangle +\big \langle n_{k-1}\big \rangle +2\big \langle n_k\big \rangle )\\ & \ \ \  \ \ \  \ \ \  \ \ \ \ \ \  \ \ \  \ \ \  \ \ \ -\gamma\delta_{l,1}(\big \langle n_{k+1}\big \rangle +\big \langle n_{k}\big \rangle )-\gamma\delta_{l,-1}(\big \langle n_{k-1}\big \rangle +\big \langle n_{k}\big \rangle ) 
\end{split}
\end{equation}
Using the translational symmetry of our system and introducing the centered and normalized pair correlation function
\begin{equation}
\label{migration}
\begin{split}
u_l (t)=\frac{ \big \langle n_i n_{i+l} \big \rangle }{ \big \langle n_i^2 \big \rangle } -1 - \frac{ \delta_{l,0} }{ \big \langle n_i \big \rangle },
\end{split}
\end{equation}
the different terms add up to
\begin{equation}
\label{Master_mean}
\begin{split}
\frac{\partial}{\partial t} u_{l} (t)  =  2 \gamma \Delta u_l (t) -2 \frac{\lambda_{SF} \overline{\nu_{SF}}}{ \langle n_{k}  \rangle } u_l (t) + \frac{\delta_{l,0}}{\langle n_{k} \rangle ^2} \big[ \lambda_F \overline {\nu_{F}(\nu_{F}-1)} \langle n_{k}\rangle + \lambda_{SF} \overline {\nu_{SF}(\nu_{SF}-1)} \big]
\end{split}
\end{equation}
Going to the limit where the cell size goes to $0$, the finite distance $x$ is obtained by considering an infinite number of cells between two points
$$ x = \lim_{\substack{h\to 0 \\ l\to \infty}} l |h|, $$
which makes it possible to introduce the spatial delta function
$$ \delta(x) = \lim_{\substack{h\to 0 \\ l\to \infty}} \frac{\delta_{l,0} }{h}, $$
as well as the continuous spatial correlation function 
$$ g(x,t) = \lim_{\substack{h\to 0 \\ l\to \infty}} \frac{u_l (t) }{h}.$$
We can finally write the equation for the pair correlation function
\begin{equation}
\label{finalsol}
\begin{split}
\Bigg[ \frac{\partial}{\partial t}-2 D \frac{\partial^2}{\partial x^2}+ 2 \frac{\lambda_{SF}^v \overline{\nu_{SF}}}{c(x,t)}\Bigg] g(x,t)=\frac{\delta (x)}{c(x,t)}\lambda_F \overline{\nu(\nu-1)}+\frac{\delta (x)}{c(x,t)^2}\lambda_{SF}^v \overline{\nu_{SF}(\nu_{SF}-1)} 
\end{split}
\end{equation}
Closely following the main lines of \cite{houchPRE1,houchPRE2}, this equation, built in the case of a 1-$d$ space can be generalized to any $d$-dimensional space by generalizing the $1$-dimensional Laplacian $\frac{\partial^2}{\partial x^2}$ to the $d$-dimensional Laplacian $\Delta ^d$ and by replacing the scalar position $x$ by the $d$-dimensional vector $r$
\begin{equation}
\label{finalsolddim}
\begin{split}
\Bigg[ \frac{\partial}{\partial t}-2 D \Delta ^d+ 2 \frac{\lambda_{SF}^v \overline{\nu_{SF}}}{c(r,t)}\Bigg] g^{dD}(r,t)=\frac{\delta (r)}{c(r,t)}\lambda_F \overline{\nu(\nu-1)}+\frac{\delta (r)}{c(r,t)^2}\lambda_{SF}^v \overline{\nu_{SF}(\nu_{SF}-1)} 
\end{split}
\end{equation}
where $g^{dD}$ refers to the spatial correlation function in dimension $D=d$.
\newline \newline
Let us consider the case of a $1$ dimensional system. The solution for negligible intrinsic sources can be retrieved from this equation by imposing $\lambda_{SF}^v=0$ and by remembering that in this case $c(x,t)=c_0$, which leads to a Gamma function solution for the correlation function
\begin{equation}
\label{clust1d}
g^{1D}( r,t ) =  \frac {1}{8} \frac { \lambda _ { F } \overline { \nu _ { F } \left( \nu _ { F } - 1 \right) } } { c_0 \pi^{3/2} D r } \: \Gamma \Big(\frac{1}{2},\frac{r^2}{8 D t} \Big)
\end{equation}
The behavior of this correlation function is rather striking since it diverges for large times, while the system (and thus the number of neutrons) is infinite. This indicates that even if fluctuations cannot kill an infinite system, spatial correlations always develop, creating holes and clusters. The patchiness in the neutron spatial distribution therefore increases with time, due to the fact that diffusions cannot compensate for the holes created by fluctuations. Whenever intrinsic sources are present, the initial condition for the neutron concentration can be set in such a way that $c_0=c_\infty= -\frac{\lambda_{SF}^v \overline{\nu_{SF}}\Lambda }{\rho}$, so that the solution to Eq.~\eqref{finalsol} can be written
\begin{equation}
\label{clustis}
\begin{split}
g^{1D} ( r,t ) & =  \frac {1}{4} \frac { \lambda _ { F } \overline { \nu _ { F } \left( \nu _ { F } - 1 \right) } } { \lambda _ { SF }^v \: \overline { \nu _ { SF }  } } \sqrt{\frac{\left|\rho \right|}{\Lambda \: D}} \: \exp \Big( -\sqrt{\frac{\left|\rho \right|}{\Lambda \: D}} \: r \Big) \\
& \times \Bigg \{ 1-\frac{1}{2}Erfc \bigg[ \frac{r}{\sqrt{8 D t}} +\sqrt{\frac{2 |\rho| t}{\Lambda}} \bigg]  \exp \Big( 2\sqrt{\frac{\left|\rho \right|}{\Lambda \: D}} \: r \Big) -\frac{1}{2}Erfc \bigg[ -\frac{r}{\sqrt{8 D t}} + \sqrt{\frac{2 |\rho| t}{\Lambda}} \bigg] \Bigg \}  
\end{split}
\end{equation}
We have assumed that $\frac{\lambda_F \overline{\nu(\nu-1)}}{c_0}>>\frac{\lambda_{SF}^v \overline{\nu_{SF}(\nu_{SF}-1)}}{c_0^2}$ since the average number of pairs produced by spontaneous fission and induced fission are similar and since $c_0/\lambda_{SF}^v>>1$.
For large times, the transient behavior of the correlation function, given by the second line of Eq.~\eqref{clustis}, is suppressed and leads to the asymptotic formula for the spatial correlation function in presence of intrinsic sources
\begin{equation}
\label{clustisasympt}
g^{1D} _ { \infty } ( r ) =  \frac {1}{4} \frac { \lambda _ { F } \overline { \nu _ { F } \left( \nu _ { F } - 1 \right) } } { \lambda _ { SF }^v \: \overline { \nu _ { SF }  } } \sqrt{\frac{\left|\rho \right|}{\Lambda \: D}} \: \exp \Big( -\sqrt{\frac{\left|\rho \right|}{\Lambda \: D}} \: r \Big)
\end{equation}
which can also be rewritten using the asymptotic neutron concentration
\begin{equation}
\label{clustisasympt2}
g^{1D} _ { \infty } ( r ) =  \frac { \lambda _ { F } \overline { \nu _ { F } \left( \nu _ { F } - 1 \right) } } { 4 \sqrt{D \: \lambda _ { SF }^v \: \overline { \nu _ { SF }  }} } \: \frac{\exp \Big( -\sqrt{\frac{\lambda _ { SF }^v \: \overline { \nu _ { SF }  }}{D \: c_\infty}} \: r \Big)}{\sqrt{c_\infty}}
\end{equation}
The qualitative behavior of this function is fundamentally different than in the case without intrinsic sources: while the fluctuations saturate, the spatial correlation function reaches an asymptotic form. It indicates that the clustering mechanism and the patchiness in the neutron spatial distribution are all the more pregnant that spontaneous fission dominates induced fission, and that diffusion is small.
\newline \newline
The previous reasoning, led in the case of a $1D$ system, can be repeated to a $2D$ system and leads to
\begin{equation}
\label{clustisasympt2_2D}
g^{2D} _ { \infty } ( r ) =  \frac { \lambda _ { F } \overline { \nu _ { F } \left( \nu _ { F } - 1 \right) } } { 4 \pi D } \: \frac{K_0 \Big( \sqrt{\frac{\lambda _ { SF }^v \: \overline { \nu _ { SF }  }}{D \: c_\infty}} \: r \Big)}{c_\infty}
\end{equation}
\newline Similarly, for $3D$ systems, the spatial correlation function can be written
\begin{equation}
\label{clustisasympt2_3D}
g^{3D} _ { \infty } ( r ) =  \frac { \lambda _ { F } \overline { \nu _ { F } \left( \nu _ { F } - 1 \right) } } { 8 \pi D } \: \frac{ \exp{\Big( -\sqrt{\frac{\lambda _ { SF }^v \: \overline { \nu _ { SF }  }}{D \: c_\infty}} \: r \Big)}}{c_\infty \: r}
\end{equation}
Finally, if a neutron detection system is designed so as to measure spatial correlations along one specific direction, $z$, the spatial correlation function measured experimentally can be retrieved by projecting Eq.~\eqref{clustisasympt2_3D} on the $z$ axis
\begin{equation}
\label{clustisasympt2_3D_proj}
g^{3D} _ { \infty } ( z ) =  \frac { \lambda _ { F } \overline { \nu _ { F } \left( \nu _ { F } - 1 \right) } } { 8 \pi D } \: \int_0 ^L dx \int_0 ^L  dy \frac{ \exp{\Big( -\sqrt{\frac{\lambda _ { SF }^v \: \overline { \nu _ { SF }  }}{D \: c_\infty}} \: \sqrt{x^2+y^2+z^2} \Big)}}{c_\infty \: \sqrt{x^2+y^2+z^2}}
\end{equation}
where $L$ stand for the typical radial dimension of the detector. If the $z$ axis follows the axial dimension of the reactor core and hence of the fuel rods, the coupling along the axial dimension is stronger than the coupling along the radial dimension, because the moderator  mainly reflects the neutron back to their initial rod (the probability that a neutron emerging from a given fuel rod enters another fuel rods is called the Dancoff factor and is roughly equal to $0.1$ in water reactors). Hence, in spite of a typical axial dimension three times larger than the radial dimension, the effective axial and radial length taking the coupling into account are such that $z<<L$. It follows that whenever the fraction between the neutron concentration arising from fissions and spontaneous fissions exceeds  $\frac{c_\infty}{\lambda _ { SF }^v}>>\frac{ \: \overline { \nu _ { SF }  }}{D}$ (verified numericaly since $D\approx10^5$ and $\overline { \nu _ { SF }  }\approx 2.5$), the integral Eq.~\eqref{clustisasympt2_3D_proj} becomes
\begin{equation}
\label{clustisasympt2_3D_proj}
g^{3D} _ { \infty } ( z ) =  \frac { \lambda _ { F } \overline { \nu _ { F } \left( \nu _ { F } - 1 \right) } } { 8 \pi D c_\infty} \Big(2 L \sinh ^{-1}(1)-\sqrt{\frac{\lambda _ { SF }^v \: \overline { \nu _ { SF }  }}{D \: c_\infty}} L^2-\frac{\pi}{2}z \Big)
\end{equation}
which can be rewritten
\begin{equation}
\label{clustisasympt2_3D_proj}
g^{3D} _ { \infty } ( z ) =  \frac { \lambda _ { F } \overline { \nu _ { F } \left( \nu _ { F } - 1 \right) } } { 8 \pi D c_\infty} \Big(2 L \sinh ^{-1}(1)-\frac{\pi}{2}z \Big)
\end{equation}
when $c_\infty>>\frac{\lambda _ { SF }^v \: \overline { \nu _ { SF }  }}{D } L^4$. Hence, the axial gradient of this function is given by
\begin{equation}
\label{clustisasympt2_3D_proj}
\frac{\partial}{\partial z} \: g^{3D} _ { \infty } =  - \frac { \lambda _ { F } \overline { \nu _ { F } \left( \nu _ { F } - 1 \right) } } { 16 D} \: \frac{1}{c_\infty}
\end{equation}
\newline
The main predictions from the modeling are the following: while the fluctuations should saturate at a level inversely proportional to the reactivity, the spatial correlation function of a 3D reactor operated at constant power measured by a one-dimensional axial detection device should be linearly decreasing along $z$ and should be decreasing as the inverse of the core power $P$. The axial gradient of this spatial correlation function should increase with $P$ as $-1/P$.

\section*{Supplementary material D: High fidelity simulations, from correlated physics to large statistics} \label{Simulations}


\subsection{From Monte Carlo simulations to numerical experiments} 
The MORET Monte Carlo neutron transport code is developed at IRSN since the beginning of the 70's. While its original goal is to support safety-criticality calculations, the head version of MORET5 has recently been upgraded to MORET6, with major functionalities targeting the ability to simulate correlated neutron physics. Adding such features required to rethink multiple parts of the code. At the basis of the correlation processes are the fission processes. Keeping the same fission model as for criticality calculations, that is integer rounding values of the number of produced neutrons, would make the neutron correlation wrong and an analog method for the fission process would thus be required. In order to use such state-of-the-art capabilities, the LLNL Fission Library~\cite{Freya} has been linked to the MORET code. Some modifications were made to this library in order to add other capabilities such as using the energy spectrum provided by MORET for the fission outgoing neutrons (needed if one wants to make the calculated $k_{eff}$ consistent between criticality and analog calculations). The LLNL Fission Library does not handle delayed neutrons and this task relies on a very simple model implemented in MORET which assumes that precursors produce at most one delayed neutron. Two initial neutron sources have been implemented: spontaneous fission and neutron generation following a user spectrum. Another important development was the so-called 'kinetic' capability, consisting in or organizing the tracking of neutrons through time bins, as the usual power iteration (tracking of neutrons through generations) and its neutron population control of criticality calculations would be meaningless in a simulation whose aim is to follow entire fission chains. Thus, two new modes have been implemented for the neutron simulations. A sequential mode allows to simulate fission chains following each other, i.e. from the first neutron to the death of all child neutrons. This mode is particularly suitable for sub-critical systems in which all fission chains die quite quickly. A chronological mode makes it possible to follow all the fission chains almost simultaneously. The simulation is divided into time steps within which the chronological order is not respected. However, this chronological order is respected from time step to time step. This type of simulation is particularly useful when it comes to simulating critical or supercritical systems that have long or infinite fission chains. During the development it was chosen to delegate the calculation of physical quantities to post-processing programs using the information extracted from the simulation. This allows both a more malleable process and avoid cluttering MORET with portions of code that are not specific to the simulation. The counterpart of this choice is the need to have outputs that are easily usable, customizable and complete. Thus all the events of the simulation can be written in a text file whose format is similar to CSV files. A multitude of filters and conditions have also been implemented in order to save only the information desired by the user and to avoid overloading these output files with useless information. Another input/output feature implemented in MORET is the ability to load and save the state of all neutrons and precursors at the beginning or the end of the simulation. Again, the chosen file format is easy to read and to modify in order to facilitate the modification or use of the file. The use of this feature for the simulation of the RCF reactor will be illustrated in the following section of this paper. All these new features will be part of the release 6 of MORET.

\subsection{Description of the simulations for design and analysis}
Among the new functionalities of MORET6, a specific attention has been devoted to the ability to simulate sub-critical and critical systems taking into account fluctuations and correlations. Tallies at the end of the simulation are not averaged following the Monte Carlo principle, but all relevant probability distribution are calculated. The only limitation to perform such high fidelity "analog" simulations are the available computing capabilities. The code has been used at all stages of the experiment, from design to data analysis.  An accurate model of the RCF and its detection equipment was used. Simulations were done in completely analog mode, meaning each neutron was tracked without using variance reduction techniques like particle splitting and Russian roulette, all fissions were accurately simulated (using the LLNL Fission Library to simulate neutron multiplicities and energies spectrum) and all secondary neutrons were simulated until each fission chain ended. However, since these simulations were designed to simulate fission chains, the simulations needed to be close to critical, the sub-reactivity worth depending on the desired thermal power of the reactor. In the simulations, the accurate model of the reactor was subcritical, and to bring the reactor critical in the simulations, the density of the fuel was increased by about 2.9\%. Then, this density was slightly adjusted in order to reach the required power.
There were also some other considerations that were taken into account, regarding delayed neutrons and the flux shape in the core. In reality, measurements were made when the reactor was critical (or near critical), meaning the proportion of delayed neutrons was near constant, however, in a normal simulation, neutrons are born at the same time (which would not be a physical representation of the measurement). Additionally, in the measurements, the reactor flux is "converged". To address both these concerns, a first set of simulations was performed, starting with an unconverged, uniform flux distribution throughout the core and with all neutrons being born at the same "time". These neutrons and all the neutrons in their fission chains were tracked until their death, with each fission location being recorded. After the first few seconds of the simulation, the flux distribution in the simulation was converged, and the locations of these fissions were used in a second set of simulations. In the second set, these fission locations were used as starting locations for neutrons. However, instead of starting all neutrons at the same time, this source neutrons were simulated as delayed neutrons, with the correct proportion of delayed neutron groups, to try to make the simulation as realistic as possible. 
Simulations were split so that each simulation had 1000 starting neutrons, and followed those neutrons and the neutrons in those fission chains until their death. Each fission event was recorded, with the time, location, event ID and father event ID (the event of the fission the neutron was born in).

\section*{Supplementary material E: Experimental setup} \label{Experiment}

To validate the predictions of the stochastic modeling an experiment was designed. This section describes the reactor used for these measurements, the detection systems used, the configurations measured, and the approach used for the experiment.

\subsection{Reactor Critical Facility}

The Walthousen Reactor Critical Facility (RCF) at Rensselaer Polytechnic Institute (RPI) is a zero power education, research, and training reactor. It was originally constructed in 1956 and fueled with HEU plates; it was converted to LEU fuel in the mid 1980's. Since then, it has been operated with 4.81 percent enriched UO$_2$ ceramic fuel with stainless steel cladding from the SPERT experiments. The fuel has an active length of 36 inches and the fuel pins are arranged in a square lattice with a pitch of 0.64 inches. The reactor is licensed to operate up to 100 Watts, but normally operates below 10 Watts. Due to the low burnup of the fuel, the fuel is essentially fresh, and can be moved by hand. The core is housed in an open pool tank which is drained of water when not in use. Coarse reactivity is provided by water height and fine reactivity is provided by control rods. The core is also surrounded by multiple uncompensated ion chambers and BF$_3$ detectors for measuring reactor power. 
\newline \newline
The RCF has a 5 Ci plutonium-beryllium (PuBe) neutron source which can be inserted near the core or stored in a paraffin block above the core. The source is inserted into or withdrawn from the reactor via an attached 1/4 inch rod using a friction drive motor. During the 2017 RCF operations, the neutron source usually giving an indication of the level of subcritical multiplication was removed from the core, ensuring that only very few of the 10$^7$ neutrons per second emitted by the source could reach the core. This was also checked by a test of the bottom-up symmetry of the neutron flux.
\newline \newline
The RCF is controlled using four control rods located on the periphery of the reactor core. These control rods are a 'flux trap' design, made of boron impregnated iron cement with a stainless steel cladding. The controls rods are 2.75 inch square tubes, with an effective length of 36 inches. The center of the control rods are hollow, and are filled with water as the control rods are inserted into the reactor. Each of the four control rods have approximately the same amount of reactivity worth. Control rods are normally inserted fully into the core until operations begin, at which point they can be lifted vertically out of the core via electric motors. The system also uses magnetic clutches, in the event of a reactor SCRAM or loss of power, the magnetic clutches are de-energized and the control rods fall into the reactor, driven by gravity. The control rods were fully withdrawn for all configurations measured in these experiments. 
\newline \newline
The reactor core is housed in a 7 foot inner diameter reactor tank made of stainless steel. When the reactor is not in operation, no water is present in the reactor tank as it is stored in a water storage tank below the reactor tank. During startup of the reactor, water is slowly pumped into the tank until the desired water height is reached. Light water from the city water supply is used as a moderator in the reactor. Due to the fact that the reactor is only ever operated to a maximum power of 15 W, no active cooling is needed. 
\newline \newline
The fuel in the RCF are Special Power Excursion Reactor Test (SPERT) F-1 pins. They are stainless steel 304 clad pins with 60 UO$_2$ pellets cylindrical each. The pellet outside diameter is 0.42 inches, cladding thickness is 0.02 inches, and cladding outside diameter is 0.466 inches. The fuels has a UO$_2$ density of 10.08 gm/cm$^3$, a $^{235}$U enrichment of 4.81 $\pm$ 0.15 weight percent, 35.2 grams of $^{235}$U per pin. The 'active length' of the fuel pins (length where fuel is present) are 36.00 inches, and the total length of the fuel pins are 41.75 inches. Inside the cladding, the fuel pellets are held in place by a spring above the fuel. Helium was used as a fill gas inside the cladding. The neutron emission rate due to spontaneous fission in $^{238}$U is $10^{-2}n/s/g$. 
\newline \newline
This type of reactor is a good choice for performing the desired measurements for several reasons. First, the ability to measure individual neutrons is required to meet the measurement goals. This therefore requires a low power level such as at a zero power reactor. Due to the absence of noticeable burnup, the fuel inside zero power reactors are typically very well characterized as compared to fuel from reactors with significant burnup. High burnup of some research reactors can also preclude entering the core for direct manipulation of experiment equipment. These considerations have been documented previously~\cite{CaSPER}. In addition, it is beneficial for the system to be physically large; while it would be beneficial to have an even larger system than the RCF, it was determined via preliminary simulations that this reactor is large enough to meet the experiment objectives.

\subsection{Detectors}
Two LANL Neutron Multiplicity $^3$He Array Detector (NoMAD) systems were placed in the reactor core, shown in Figure 1. Each NoMAD system contain 15 $^3$He detectors arranged into three rows, and each detector is 15 inches in length \cite{SCRaP}; additional information on the $^3$He tubes is given in Table~\ref{tab:NoMAD_He3}. In between the detectors is a polyethylene moderator. Due to the sensitive electronics of the detectors, they were placed in waterproof aluminum enclosures. The detector enclosures containing the NoMAD systems were placed on an aluminum stand and stacked on top of one another so that the detector tubes laid horizontally. This allowed for measurement of the flux as a function of the axial position. The height was positioned such that the detectors were centered axially with the fuel pins. Radially, the distance from the center of the reactor to the face of the detector enclosures was 55~cm as inicated by Figure 1. The NoMAD systems have been used for many neutron multiplicity measurements; most of these experiments have taken place at the National Criticality Experiments Research Center (NCERC) \cite{SCRaP_Part2,SCRaP,NDSE,BeRP_GA,NoMAD+OS}. The NoMAD was previously used for subcritical measurements at the RCF \cite{CaSPER}; lessons learned from that measurement campaign were utilized in the design of this experiment.

\begin{longtable}{|l|l|}
\caption{NoMAD $^3$He Tube Information}
\label{tab:NoMAD_He3}\\
\hline 
\hline
Manufacturer & Reuter-Stokes \\ \hline
Model Number & RS-P4-0815-103 \\ \hline
Body Material & Aluminum 1100 \\ \hline
External Diameter & 1.00 inch \\ \hline
Thickness & 1/32 inch \\ \hline
Height (including cladding) & 41.6 cm \\ \hline
$^3$He Pressure & 150 psia \\ \hline
Active Length & 15.0 inch \\ \hline
\hline
\end{longtable} 

\noindent
Also placed in the core were four small $^3$He detectors, shown in Figure 1. The detectors are manufactured by Reuter-Stokes (RS-P4-0203-201), with a 0.25 inch diameter, an active length of 2.99
inches, and a $^3He$ pressure of 40~atm. These detectors fit into a single fuel pin location in the core (an empty fuel pin with an open top was manufactured so that these detectors could be placed in the core). These detectors were placed at four heights so that the in core flux could also be measured axially (see Figure 1). These $^3$He detectors have been used in many Rossi-$\alpha$ measurements at NCERC~\cite{NCERC_Rossi,Pb+Rossi,Krusty+Rossi,Foils+Rossi,Zeus+Rossi,NDSE,HuaRossi}.
\newline \newline
Both the NoMAD and small $^3$He detector systems record list-mode data. The raw data includes a list of times (to the nearest 100 ns) and channel associated with every event that was recorded. An overall view of all the detectors positioned inside the RCF is given in Figure1, which contains both pictures and MORET6 modeling views of all the components of this setup.
\newline \newline
Inside the reactor, surrounding the core are five permanent detectors used to measuring reactor power. Two of the detectors are BF$_3$ detectors and are used as startup detectors in pulse mode (labeled Startup Channels A and B). Once a high enough power is reached, three additional uncompensated ion detectors are used in current mode. Two of these detectors are displayed in the control room on linear scales (labeled LP1 and LP2), and one is displayed on a log scale (PP2). Reactor period is also automatically calculated and displayed based on the detector response from PP2. 
\newline \newline

\subsection{Runs and experiments performed}
The experiments were conducted from August 14-18, 2017. The following personnel conducted the measurements: Rian Bahran, Eric Dumonteil, Jesson Hutchinson, George McKenzie, Alex McSpaden, Wilfried Monange, Mark Nelson, and Nicholas Thompson; licensed RPI Reactor staff operated the reactor facility during the course of the experimental operation and are listed in the acknowledgements below. Table~\ref{tab:ConfigurationsFull} shows a list of all measured configurations. Some of the configurations in Table~\ref{tab:ConfigurationsFull} were subcritcal or supercritical. Some of these configurations even included elements which were changing; this included water filling, control rod position changing, and SCRAMs. Some of these measurements may be used for analysis such as rod drop technique, but others are not necessarily for analysis and were used to ensure detector functionality.
\newline \newline
As shown in Table~\ref{tab:ConfigurationsCritical}, 11 different power levels were measured. All of these configurations had 335 fuel pins, 4 small $^3$He tubes present, and were critical. These are the specific configurations which are utilized for the results shown in Section~\ref{subsec:dettostoch}.

\begin{longtable}{|l|l|l|l|l|l|}
\caption{RCF Full List of Measured Configurations}
\label{tab:ConfigurationsFull}\\
\hline 
\hline
Date & \# Fuel Pins & Criticality State & Power & \# Small $^3$He Tubes & Water Present \\
\hline
8/15/2017 & 334 & Supercritical & Changing & 0 & Yes \\ \hline
8/15/2017 & 333 & Supercritical & Changing & 0 & Yes \\ \hline
8/15/2017 & 334 & Critical & ? & 0 & Yes \\ \hline
8/15/2017 & 335 & Supercritical & Changing & 1 & Yes \\ \hline
8/15/2017 & 335 & Supercritical & Changing & 2 & Yes \\ \hline
8/15/2017 & 335 & Supercritical & Changing & 3 & Yes \\ \hline
8/15/2017 & 335 & Supercritical & Changing & 4 & Yes \\ \hline
8/15/2017 & 335 & Subcritical & - & 4 & Yes \\ \hline
8/15/2017 & 335 & Subcritical & - & 4 & Yes \\ \hline
8/15/2017 & 335 & Critical & 1 W & 4 & Yes \\ \hline
8/15/2017 & 335 & Subcritical & - & 4 & Yes \\ \hline
8/15/2017 & 335 & Critical & 100 mW & 4 & Yes \\ \hline
8/15/2017 & 335 & Subcritical & - & 4 & Yes \\ \hline
8/15/2017 & 335 & Critical & 10 mW & 4 & Yes \\ \hline
8/15/2017 & 335 & Subcritical & - & 4 & Yes \\ \hline
8/16/2017 & 335 & Subcritical & - & 4 & No \\ \hline
8/16/2017 & 335 & Subcritical & - & 4 & No \\ \hline
8/16/2017 & 335 & Subcritical & - & 4 & Filling \\ \hline
8/16/2017 & 335 & Subcritical & - & 4 & Yes \\ \hline
8/16/2017 & 335 & Subcritical & - & 4 & Yes \\ \hline
8/16/2017 & 335 & Subcritical & - & 4 & Yes \\ \hline
8/16/2017 & 335 & Critical & 1.15 mW & 4 & Yes \\ \hline
8/16/2017 & 335 & Supercritical & Changing & 4 & Yes \\ \hline
8/16/2017 & 335 & Critical & 4.83 mW & 4 & Yes \\ \hline
8/16/2017 & 335 & Subcritical & - & 4 & Yes \\ \hline
8/16/2017 & 335 & Critical & 1.8 mW & 4 & Yes \\ \hline
8/16/2017 & 335 & Supercritical & Changing & 4 & Yes \\ \hline
8/16/2017 & 335 & Critical & 1.64 mW & 4 & Yes \\ \hline
8/16/2017 & 335 & Supercritical & Changing & 4 & Yes \\ \hline
8/16/2017 & 335 & Critical & 100 mW & 4 & Yes \\ \hline
8/16/2017 & 335 & Subcritical & - & 4 & Yes \\ \hline
8/17/2017 &  & Subcritical & - & 4 & Filling \\ \hline
8/17/2017 & 335 & Subcritical & - & 4 & Yes \\ \hline
8/17/2017 & 335 & Critical & 10 mW & 4 & Yes \\ \hline
8/17/2017 & 335 & Supercritical & Changing & 4 & Yes \\ \hline
8/17/2017 & 335 & Critical & 50 mW & 4 & Yes \\ \hline
8/17/2017 & 335 & Critical & 500 mW & 4 & Yes \\ \hline
8/17/2017 & 335 & Critical & 500 mW & 4 & Yes \\ \hline
8/17/2017 & 335 & Critical & 1 W & 4 & Yes \\ \hline
8/17/2017 & 335 & Subcritical & - & 4 & Yes \\ \hline
8/17/2017 & 335 & Subcritical & - & 4 & No \\ \hline
8/17/2017 & 335 & Subcritical & - & 4 & Yes \\ \hline
8/17/2017 & 335 & Critical & 0.5 mW & 4 & Yes \\ \hline
8/17/2017 & 335 & Critical & 5 mW & 4 & Yes \\ \hline
8/17/2017 & 335 & Critical & 5 mW & 4 & Yes \\ \hline
\hline
\end{longtable}

\begin{longtable}{|l|l|}
\caption{RCF List of Critical Configurations in Order of Increasing Power Level}
\label{tab:ConfigurationsCritical}\\
\hline 
\hline
Date & Power \\ \hline
8/17/2017 & 0.5 mW \\ \hline
8/16/2017 & 1.15 mW \\ \hline
8/16/2017 & 1.64 mW \\ \hline
8/16/2017 & 1.8 mW \\ \hline
8/16/2017 & 4.83 mW \\ \hline
8/17/2017 & 5 mW \\ \hline
8/17/2017 & 5 mW \\ \hline
8/15/2017 & 10 mW \\ \hline
8/17/2017 & 10 mW \\ \hline
8/17/2017 & 50 mW \\ \hline
8/15/2017 & 100 mW \\ \hline
8/16/2017 & 100 mW \\ \hline
8/17/2017 & 500 mW \\ \hline
8/17/2017 & 500 mW \\ \hline
8/15/2017 & 1 W \\ \hline
8/17/2017 & 1 W \\ \hline
\hline
\hline
\end{longtable}

\noindent
In order to bring the reactor critical with an unknown configuration, multiple measurements and tests were performed. In particular, 1/M measurements were taken as a function of control rod height to determine expected critical bank rod position and control rod worths (both differential and integral). 
\newline \newline
The much larger NoMAD systems did not change reactivity considerably because of the distance away from the core and since they primarily consist of polyethylene which is very similar to water. However, due to the large amount of negative reactivity that each in core $^3$He detector added to the reactor, the detectors had to be added one at a time, and fuel also had to be added to bring the reactor back to critical as shown in Table~\ref{tab:ConfigurationsFull}. Each time a change was made to the system, that core configuration needed to be qualified, so a number of tests needed to be performed (for example, SCRAM tests, measurements of the reactivity of the most reactive pin, and measurements of the shutdown reactivity). 

\subsection{Sources of systematic errors}
There are a number of sources of systematic error that should be pointed out. First is the exact location of the small $^3$He detectors. Because multiple small $^3$He detectors were placed into the empty pin, it was not possible to definitively measure the exact locations of each detector. Determining the error in the detector locations and the impacts on the data will be discussed in the analysis section. 
The reactor normally operates at a "much higher" power than most of these measurements were taken at (usually between 1-10 Watts), and there was some uncertainty about the exact reactor power during measurements. This will be discussed later in the analysis section. 
\newline \newline
Lastly, there is some uncertainty with the control rod heights measured. The control rod heights are measured using an optical encoder connected via gears to the shaft of the control rod. However, in rare cases, the gears for the optical encoder can occasionally slip, resulting in faulty readings for control rod heights. This can be seen pretty easily in the control panel. The uncertainty in the control rod heights and the impacts of this on the results will also be discussed further in the analysis section.  

\subsection{Power calibration}
Because the RCF is a low power reactor, reactor power cannot be measured via the heat produced from fission. The typical approach to power calibration of the RCF is to irradiate gold foils. Small gold foils are placed on three locations along the center fuel pin of the reactor, and the reactor is brought to critical at a set indicated power (eg. 0.1 Watts) for approximately 30 minutes\footnote{One important note: the power in the control room is measured using compensated ion chambers (operated in an uncompensated mode) in current mode. This current is converted into power using the conversion ratio from the previous power calibration.}. After the irradiation, the reactor is scrammed and the gold foils are removed from the fuel pin. These gold foils are then measured using a gamma detector to determine radioactivity. The radioactivity of the gold foils is then used to calculate the flux and power that the reactor was operating at, using a conversion between detector counts and reactor flux which was simulated using MCNP\textsuperscript{\textregistered}6\footnote{MCNP\textsuperscript{\textregistered} and Monte Carlo N-Particle\textsuperscript{\textregistered} are registered trademarks owned by Triad National Security, LLC, manager and operator of Los Alamos National Laboratory. Any third party use of such registered marks should be properly attributed to Triad National Security, LLC, including the use of the designation as appropriate. For the purposes of visual clarity, the registered trademark symbol is assumed for all references to MCNP within the remainder of this paper.}
~\cite{mcnp}. The calculated power is then compared against the indicated power, and the conversion between current and indicated power is adjusted if necessary. 
During each step of this process there are potential sources of error. There is a small amount of error in the recorded detector signals, the sum of which is needed for the calibration. There is a small amount of error in the size and mass of the gold foils. There is some error in the vertical position of the gold foils on the fuel pin. After the irradiation, there is some error in the measurement, both because the detector efficiency may not be calibrated perfectly and because the ROI for the gold peak may be set differently in each power calibration. Finally, there are also uncertainties and a potential bias in the MCNP simulation due to model approximations, nuclear data uncertainties, and statistical uncertainties.
\newline \newline
The RCF detector responses were analyzed and count rates were translated into reactor power, based on the power calibration for that core (as adding fuel pins and detectors can change the power calibration). At powers near 0.1 watt or above, the three RCF detectors are in complete agreement. However, below 0.1 watt, differences between the three detectors (LP1, LP2, PP2) begin to show up, due to the different sensitivities and background noise of the detectors. LP2 is the most sensitive detector, followed by PP2 and then LP1. Because of this, at very low powers, the three detectors will begin to disagree, with the power calculated from the LP1 count rate being much higher than that of PP2, and PP2 having a higher calculated power than LP2. 
Additionally, these detectors are compensated ion chambers, operated in an uncompensated mode, and some of the detected counts are from gamma rays. This introduces an additional problem, since after operating at a high power, if the reactor power is lowered by an order of magnitude, there is still an additional gamma background which is seen by the detectors and which decays over time. 
As explained previously, each NoMAD system is made up of an array of 15 long $^3$He detectors, and two of these systems were stacked on top of each other, with the tubes in a horizontal orientation. These $^3$He detectors are insensitive to gamma background and are more sensitive than the LP2 detector, making them much better suited for determining reactor power at low powers. At powers above 0.5 Watts, some of the detector tubes suffered significant amounts detector deadtime. At very high powers (1 Watt and above), system deadtime also became an issue. But for powers 0.1 Watts and below, power calculated from NoMAD detector counts agreed with the calculated power from the RCF detectors.  
The NoMAD detector count rates were calibrated to reactor power, with each detector tube calibrated individually and deadtime correction performed. However, when looking at the ratios of detector counts in each detector tube, the ratios changed between measurements. This effect was analyzed and it was determined that changes in the reactor configuration (eg. adding fuel pins or additional small $^3$He detectors) changed the detector responses relative to the other detectors. Additionally, a second effect was seen, where changes in control rod position impacted the count rates for detector tubes near the top of the reactor core (control rod shadowing). This effect is very clear, and even small changes in control rod positions change the ratios of count rates between detector tubes at the top of the core vs. the rest of the detector tubes. However, this effect is negligible for detector tubes in the bottom half of the core, since the control rods are inserted from the top of the core. Therefore, the detector tubes in the top half of the core were ignored when calculating power. 
In summary, power was calibrated using the gold foil irradiation technique, giving a detector calibration for each detector. The LP2 detect, the most sensitive detector, was used to calibrate the NoMAD detectors. For low powers (under 0.1 Watts) the detectors in the bottom NoMAD system were used to calculate power. In all other cases, a determination was made based on deadtime and LP2 response as to which system was more accurate for calculating power. 

\end{addendum}

\newpage
\section*{Figures}

\newpage

\newpage

\newpage

\newpage

\newpage

\end{document}